%% file: captcha-acm-csur.tex
\documentclass[acmsmall,screen,authordraft,nonacm,review=false,timestamp=false]{acmart}

\AtBeginDocument{%
  \providecommand\BibTeX{{%
    \normalfont B\kern-0.5em{\scshape i\kern-0.25em b}\kern-0.8em\TeX}}}

\setcopyright{none}
\acmJournal{CSUR}
\acmVolume{0}
\acmNumber{0}
\acmArticle{0}
\acmMonth{0}

\newcommand \imgW {3cm}

\newcommand \imgL {2cm}
\newcommand \imgT {0.8cm}

\usepackage{placeins}
\usepackage{ltablex,booktabs}
\usepackage{multirow}
\usepackage{longtable}
\usepackage{afterpage}

\newboolean{showcomments}
\setboolean{showcomments}{true}         %\setboolean{showcomments}{false} 

\ifthenelse{\boolean{showcomments}}
  {\newcommand{\nb}[2]{
  \fbox{\bfseries\sffamily\scriptsize#1}
     {\sf\small$\blacktriangleright$\textit{\textcolor{red}{#2}}$\blacktriangleleft$}
   }
  }
  {\newcommand{\nb}[2]{}
   
  }

\graphicspath{ {./textimages/} }
\graphicspath{ {./otherimages/} }

\begin{document}

\title{Gotta CAPTCHA 'Em All: A Survey of Twenty years of the Human-or-Computer Dilemma}


\author{Meriem Guerar}
\orcid{1234-5678-9012}
\authornotemark[1]
\email{meriem.guerar@dibris.unige.it}
\affiliation{%
  \institution{DIBRIS, University of Genoa}
  \city{Genoa}
  \country{Italy}
}

\author{Luca Verderame}
\orcid{0000-0001-7155-7429}
\email{luca.verderame@dibris.unige.it}
\affiliation{%
  \institution{DIBRIS, University of Genoa}
  \city{Genoa}
  \country{Italy}
}

\author{Mauro Migliardi}
\orcid{0000-0001-7155-7429}
\email{mauro.migliardi@unipd.it}
\affiliation{%
  \institution{Department of Electronic Engineering, University of Padua}
  \city{Padua}
  \country{Italy}
}

\author{Francesco Palmieri}
\orcid{0000-0001-7155-7429}
\email{fpalmieri@unisa.it}
\affiliation{%
  \institution{Department of Computer Science, University of Salerno}
  \city{Salerno}
  \country{Italy}
}

\author{Alessio Merlo}
\orcid{0000-0001-7155-7429}
\email{alessio.merlo@dibris.unige.it}
\affiliation{%
  \institution{DIBRIS, University of Genoa}
  \city{Genoa}
  \country{Italy}
}

\renewcommand{\shortauthors}{Guerar et al.}

\begin{abstract}
A recent study has found that malicious bots generated nearly a quarter of overall website traffic in 2019
%, and that 16.6\% of the traffic is coming from mobile devices 
\cite{ref137}. These malicious bots perform activities such as price and content scraping, account creation and takeover, credit card fraud, denial of service, etc. Thus, they represent a serious threat to all businesses in general, but are especially troublesome for e-commerce, travel and financial services.
One of the most common defense mechanisms against bots abusing online services is the introduction of Completely Automated Public Turing test to tell Computers and Humans Apart (CAPTCHA), so it is extremely important to understand  which CAPTCHA schemes have been designed and their actual effectiveness against the ever-evolving bots. To this end, this work provides an overview of the current state-of-the-art in the field of CAPTCHA schemes and defines a new classification that includes all the emerging schemes. In addition, for each identified CAPTCHA category, the most successful attack methods are summarized by also describing how CAPTCHA schemes evolved to resist bot attacks, and discussing the limitations of different CAPTCHA schemes from the security, usability and compatibility point of view. Finally, an assessment of the open issues, challenges, and opportunities for further study is provided, paving the road toward the design of the next-generation secure and user-friendly CAPTCHA schemes.
\end{abstract}

\begin{CCSXML}
<ccs2012>
   <concept>
       <concept_id>10002978.10002991.10002992</concept_id>
       <concept_desc>Security and privacy~Authentication</concept_desc>
       <concept_significance>500</concept_significance>
       </concept>
   <concept>
       <concept_id>10002978.10002991.10002992.10011618</concept_id>
       <concept_desc>Security and privacy~Graphical / visual passwords</concept_desc>
       <concept_significance>500</concept_significance>
       </concept>
 </ccs2012>
\end{CCSXML}

\ccsdesc[500]{Security and privacy~Authentication}
\ccsdesc[500]{Security and privacy~Graphical / visual passwords}

\keywords{CAPTCHA, Bot, CAPTCHA Type, Security, Text CAPTCHA, Image CAPTCHA, Behavior CAPTCHA, Sensor CAPTCHA.}

\maketitle

\section{Introduction}\label{sec:introduction}

A \emph{Completely Automated Public Turing tests to tell Computers and Humans Apart} (CAPTCHA) is, as the name suggests, a challenge-response test used to distinguish between genuine human users and automated computer programs. CAPTCHAs are commonly used to prevent abuses of online services such as registering thousands of free accounts, obtaining tickets for resale, spreading spam emails, taking over accounts by using brute force \cite{ref4}, or perform credential stuffing attacks \cite{ref3}.

The idea of using a CAPTCHA to check whether the users who are making requests to a web service are humans goes back to 1996 \cite{ref124}. 
A year later, AltaVista developed the first practical example of a CAPTCHA scheme, which was based on the inability of Optical Character Recognition (OCR) software to recognize a distorted text \cite{ref125}. 

In 2000, Von Ahn et al. \cite{ref123}, \cite{von2003captcha} introduced several practical proposals for designing CAPTCHA schemes based on \emph{hard Artificial Intelligence (AI) problems}, i.e., challenges that most humans can solve easily, but computer programs cannot pass. 

Most CAPTCHA schemes proposed in the literature follow such an approach and exploit different elements such as character recognition, image understanding, and speech recognition to create challenges that successfully block automated bots.
However, the recent advancement of AI in general and Computer Vision (CV) in particular has made automated programs significantly better at solving such tests. As a result, almost all of the traditional CAPTCHA schemes have been broken as demonstrated in \cite{ref20}, \cite{ref130}, \cite{ref149} and \cite{ref41}.

Furthermore, in contrast to Von Ahn et al. expectations, not all the attacks proposed in the literature attempt to solve the underlying AI problem on which these CAPTCHAs are based in order to break them. Some of them, instead, try to circumvent the AI problem by leveraging the weaknesses in the design of a particular CAPTCHA scheme \cite{ref122}, \cite{ref58}, \cite{ref48}. These kinds of attacks are known as side-channel attacks. 

Over time, designing effective and user-friendly CAPTCHA schemes based on hard AI problems has become very challenging. This has led to the emergence of a new generation of  schemes based on behavioral analysis and sensor readings.

In 2014 Google announced that today's Artificial Intelligence technology can solve even the most difficult variant of distorted text at 99.8\% accuracy \cite{ref71} and moved to a CAPTCHA scheme based on behavioral analysis which is considered the dominant CAPTCHA scheme in the market today. In the academic world, many works have shown the vulnerability of the traditional CAPTCHA schemes, nevertheless, many researchers still aim at breaking traditional CAPTCHA schemes and evaluating their security and usability \cite{ref145}, \cite{ref146}, \cite{ref147}, \cite{ref148}, ignoring the emerging CAPTCHA schemes that have not been broken yet. 
%In addition, many others do not consider the new threats such as fourth generation bots and human solver relay attacks in their CAPTCHA design.
Still, recent works in the literature do not consider these new CAPTCHA schemes neither in their review nor in their security evaluation \cite{ref131},\cite{ref134}, \cite{ref27}.

%In the academic world, many researchers still aim at breaking traditional CAPTCHA schemes and evaluate their security and usability; however, in 2014 Google announced that today's Artificial Intelligence technology can solve even the most difficult variant of distorted text at 99.8\% accuracy and many academic works show the vulnerability of the traditional CAPTCHA schemes. they moved to behavior-based CAPTCHA schemes and they designed what accounts to the dominant CAPTCHA scheme in the market today.

\paragraph*{\textbf{Contribution}} Different from the existing CAPTCHA surveys (e.g., \cite{ref133}, \cite{ref160}, \cite{ref131},\cite{ref134}), in this work, we present an up-to-date comprehensive CAPTCHA survey that includes both the traditional CAPTCHA schemes and the new generation ones, such as those based on behavior and sensor readings. Then, we propose a novel classification of the existing CAPTCHA literature from 2000 to 2020 based on \emph{ten different groups} (i.e., Text-based, Image-based, Audio-based, Video-based, Game-based, Slider-based, Math-based, Behavior-based, Sensor-based, and CAPTCHA for liveliness detection). To the best of our knowledge, this is the first survey that reviews behavioral-based, sensor-based CAPTCHAs, and CAPTCHA designed for liveliness detection in authentication methods. 
Furthermore, we survey and analyze all the literature regarding the security evaluation of the existing CAPTCHA schemes and the proposed techniques to break them, showing the weaknesses of the different categories of CAPTCHA schemes. 
This work also allows us to build a timeline for the security of 77 CAPTCHA schemes illustrating the creation and breaking year along with the breaking percentage. Besides showing the evolution of CAPTCHA over two decades, this timeline provides a clear view of the broken CAPTCHA mechanisms and the ones that are worth further investigation. In addition, it elucidates the new design trends in CAPTCHA schemes.

Finally, we discuss the evolution of CAPTCHA schemes in terms of new design trends, their security, and their user-friendliness; moreover, we illustrate the open issues, the challenges, and the opportunities for further study, drawing a roadmap for the design of the next generation of secure and user-friendly CAPTCHA schemes.

\paragraph*{\textbf{Structure}} The rest of this paper is organized as follows. In Section \ref{sec:classification}, we introduce a comprehensive classification of conventional and recent emerging CAPTCHA schemes. In Section \ref{sec:attacks}, we revise the main attacks against the CAPTCHA schemes described in section \ref{sec:classification}. In Section \ref{sec:discussion}, we provide a discussion on the current state-of-the-art of CAPTCHA, highlighting the CAPTCHA evolution and the limitations of each CAPTCHA design from different standpoints. Section \ref{sec:challenge}, discusses open issues, challenges and opportunities for future work.
%In Section \ref{sec:recommendations}, we provide requirements and recommendations for designing the next generation CAPTCHA.
Finally, in Section \ref{sec:conclusion}, we draw some conclusions from all the analyses and comparisons performed.

\section{CAPTCHA classification}\label{sec:classification}

The traditional classification of CAPTCHA in the literature defines six categories, namely text-based, image-based, audio-based, video-based, math-based and game-based CAPTCHA \cite{ref132}, \cite{ref133}. However, we consider this classification incomplete because it does not cover the new emerging CAPTCHA schemes. As an example, the most widely adopted CAPTCHA schemes today do not fall into this classification (e.g., reCAPTCHA V2 and Geetest).  Nevertheless, even the most recent surveys in the literature adopt this incomplete classification to review and evaluate the security of the existing CAPTCHA schemes \cite{ref131},\cite{ref134}, \cite{ref27}. This discrepancy between the relevant literature and the actual state of the art motivated us to propose a more comprehensive classification capable of capturing the new emerging CAPTCHA schemes. % This is important to highlight the development and the new directions of CAPTCHA design. 
We argue that current CAPTCHA schemes can be divided into 10 categories, i.e., \emph{Text-based, Image-based, Audio-based, Video-based, Game-based, Slider-based, Behavior-based, Sensor-based, and CAPTCHAs for liveliness detection in authentication methods}.

%We argue that current CAPTCHA schemes can be divided into the following categories,

%\begin{itemize}

%\item Text-based CAPTCHA
%\item Image-based CAPTCHA
%\item Audio-based CAPTCHA
%\item Video-based CAPTCHA
%\item Game-based CAPTCHA
%\item Slider-based CAPTCHA
%\item Math-based CAPTCHA
%\item Behavior-based CAPTCHA
%\item Sensor-based CAPTCHA
%\item CAPTCHAs for liveliness detection in authentication methods
%\end{itemize}

It is important to mention that the new CAPTCHA schemes that involve a traditional challenge/response test belong to the old category as well; yet, in order to highlight the development and the new directions in CAPTCHA design, we will focus on the new added mechanisms. 

\subsection{Text-based CAPTCHAs}\label{sec:text captcha}

% Text-based CAPTCHAs are the most popular form of CAPTCHA where distorted texts (i.e., English world or a sequence of characters) are shown as CAPTCHA images. The user is asked to input the text appearing in the image in order to pass the test. The idea behind using the distorted text is that it is assumed to be easy for human users to read it, yet difficult for bots to recognize it using Optical Character Recognition (OCR) techniques. Since the users interaction required to solve the CAPTCHA (i.e., type the text) is the same in almost all text-based CAPTCHAs, the proposed classification is based on the representation of the text in the challenge displayed to the user. Hence, the existing text-based CAPTCHAs in the literature can be classified into the following three categories: 1) 2D text-based CAPTCHA, 2) 3D text CAPTCHA and 3) Animated text-based CAPTCHA. 

Text-based CAPTCHAs are the most popular form of CAPTCHA; in these schemes a text (e.g., a sequence of random characters or words) is distorted and displayed to the user as an image. When words are used, language dependency represents a major limitation of this kind of CAPTCHA schemes. Then, the user is asked to input the text appearing in the image to pass the test. The underlying assumption is that humans can read the distorted text easily, but this is hard for bots using Optical Character Recognition (OCR) techniques.

%The distortion effects and the representation of the text as an image \footnote{isn't OCR ALWAYS applied to an image? If this is the case, the fact that the CAPTCHA is an image is irrelevant from the OCR point of view} enable Text-based CAPTCHAs to interfere with Optical Character Recognition (OCR) used by bots to solve the test.

Since the interaction required to solve the CAPTCHA (i.e., the input of a text) is the same in almost all text-based CAPTCHAs, we classified the variation of Text-based CAPTCHAs according to the different representation of the text of the challenge. Hence, we identified three sub-categories: 1) 2D text-based, 2) 3D text-based, and 3) Animated text-based. Table ~\ref{table:textCAPTCHA} gathers all the considered text-based CAPTCHA schemes, a relevant graphical sample, and a detailed description of the challenge.

\subsubsection{2D text-based CAPTCHA}\label{sec:2D captcha}

The 2D text-based CAPTCHA scheme was initially developed by Andrei Broder and his colleagues at the DEC Systems Research Center in 1997. In the same year, the AltaVista website used such a method to block bots trying to influence the rank of a set of sites on the AltaVista search engine \cite{ref162}. 

In 2000, Von Ahn and Blum, in collaboration with Yahoo, developed \textbf{Gimpy CAPTCHA} and \textbf{EZ-Gimpy} \cite{ref9} to prevent spammers from posting malicious advertisements in the chat rooms and to ensure that free accounts were granted only to real individuals. 
The challenge of the Gimpy CAPTCHA scheme consists of typing correctly at least three out of seven words randomly selected from a dictionary. 

EZ-Gimpy is a simplified version of Gimpy showing only a single random word selected from the dictionary.
However, the word is rendered to an image using different fonts, background grids and gradients. Furthermore, the image is altered by using blurring, noise and distortion effects on letters.

In 2003, Monica Chew and Henry Baird proposed \textbf{BaffleText} \cite{ref12}, a text-based CAPTCHA scheme that adopts pseudo-random but pronounceable words along with some masking techniques aiming at preventing the use of OCR software.

In 2010, the popular website for sharing and uploading files \textbf{(Megaupload.com)} designed a CAPTCHA scheme based on a new segmentation-resistant mechanism different to that used by \textbf{Microsoft}, \textbf{Google} and \textbf{Yahoo}. This new mechanism relies on the combination of overlapping characters and the “Gestalt Perception” principle which is used to hide some contents of the characters where they connect to each other. The Gestalt Perception principle suggests that humans can reconstruct individual characters mentally, while this task is still difficult for computer programs.

The most widely deployed form of text-based CAPTCHA is the first version of \textbf{ReCAPTCHA} \cite{ref19}, which had the two-fold aim of protecting websites from bot attacks and digitize old books. 
The challenge consists of recognizing two distorted words scanned from old books, one known by the algorithm and one that OCR programs have failed to identify. The challenge is successfully passed if the user correctly recognizes and types the known word. Besides, if the challenge is passed, the algorithm assumes that the user recognized also the second unknown word.

To improve the usability of text-based CAPTCHAs, Chow et al. \cite{ref21} introduced the idea of \textbf{clickable CAPTCHA}. Their approach consists of combining multiple textual CAPTCHA challenges into a grid of clickable CAPTCHAs (e.g., a 3-by-4 grid). The user has to click on the grid elements that match the challenge requirement. 
For instance, the challenge can be the identification of English words among non-English words in the grid. Obviously, such a challenge has language dependencies. 

In contrast to traditional CAPTCHA schemes that use machine-printed text, authors in \cite{ref22,ref23} proposed \textbf{Handwritten CAPTCHAs} that use as challenges synthetic handwritten text images, already known to fool OCR software.

%%%%%%%%%%%%%%%%%%%%%%% TABLE 1 GOES HERE %%%%%%%%%%%%%%%%%%%%%%%%

\input{tables/table1}

\subsubsection{3D text-based CAPTCHA}\label{sec:3D captcha}

3D text-based CAPTCHA schemes exploit the fact that human beings can easily recognize sequences of 3D characters while bot programs cannot; thus, they represent an advancement in comparison to the 2D text-based CAPTCHA schemes.

One of the first proposals is the \textbf{Teabag 3D} designed by the OCR Research Team \cite{ref7} to identify the weaknesses of 2D text-based CAPTCHA schemes and propose a novel -- and more secure -- CAPTCHA scheme. Teabag 3D consists of an image with a 3D pattern that contains textual characters (as shown in Table ~\ref{table:textCAPTCHA} ). Thanks to the new CAPTCHA scheme, the authors demonstrated that humans could easily recognize the 3D text and, at the same time, automated systems failed in the recognition task.

Similarly, \textbf{Super CAPTCHA} \cite{ref8} and \textbf{3DCAPTCHA} \cite{ref26} are 3D text-based CAPTCHA schemes that were based on those same assumptions and used on several websites.
For instance, Super CAPTCHA is also available as a plug-in for WordPress.org since 2013 \footnote{https://wordpress.org/plugins/super-capcha/\#description}.

Imsamai and Phimoltares \cite{ref24} introduced the \textbf{3D CAPTCHA} scheme by rendering a sequence of 3D alphanumeric characters and applying a set of different effects to trick automated recognition systems. Those effects include text rotation, text overlapping, noise addition, scaling, font variation, special characters and different background textures. 

Recently, Suzi et al.\cite{ref28} introduced a new type of 3D text-based CAPTCHA, called \textbf{DotCHA}. The challenge consists of 3D letters composed of several small spheres. Each character is twisted around a horizontal axis so that each letter is readable at a different rotation angle. Thus, the user needs to rotate the 3D text model multiple times to identify all the letters. 
From the usability point of view, DotCHA adds an additional task (i.e., the rotation of the model multiple times) in comparison to the traditional text-based CAPTCHAs that require only the input of the text to solve the challenge.

\subsubsection{Animated text-based CAPTCHA}\label{sec:anim captcha}

%Animated CAPTCHAs are one of the design variant of Text-based CAPTCHAs that incorporate the time dimension into the challenge through animation. It is based on the assumption that humans can easily process the moving information, while computers will have difficulty to extract this information. 

%One of the first proposals of animated CAPTCHA has been introduced by Fischer and Herfet \cite{ref29} in 2006. Their CAPTCHA is based on the idea of projecting distorted text onto a deforming animated surface. In 2009, Naumann et al. \cite{ref30} introduced an animated CAPTCHA based on the perception that the human visual system tends to group different entities that move together. Authors exploit this specific capability of the human visual system to develop a new CAPTCHA that shows letters superimposed over a noisy background. The users will be able to distinguish the text from the background when the letters are moving. Similarly, Cui et al. \cite{ref31} proposed an animated CAPTCHA where the user can get the right characters shown in the animation only when they are moving. In addition, they introduced the "zero knowledge per frame" principle which ensures that each frame of the animation would not leak enough information that solve the CAPTCHA.

Animated CAPTCHAs extend text-based schemes by introducing the time dimension. In details, these CAPTCHA schemes animate the textual content in the challenge in a short clip, thus complicating the extraction task for automated systems.

One of the first proposals of animated CAPTCHA has been introduced by Fischer and Herfet \cite{ref29} in 2006. Their CAPTCHA scheme is based on the idea of projecting the text onto a deforming animated surface. 
In 2009, Naumann et al. \cite{ref30} introduced an animated CAPTCHA based on the perception that the human ocular system tends to group different entities that move together. Hence, the authors developed a new CAPTCHA scheme that shows letters superimposed over a noisy background. The users are able to distinguish the text from the background when the letters are moving. 

Similarly, Cui et al. \cite{ref31} proposed an animated CAPTCHA where the user can get the right characters shown in the animation only when they are moving. They also introduced the ``zero-knowledge per frame'' principle, which ensures that each frame of the animation does not leak enough information to solve the CAPTCHA challenge.

Besides the CAPTCHA schemes proposed by the scientific community, there are a set of solutions offered either by specific websites or by CAPTCHA service providers. 

For instance, the Creo Group \cite{ref32} introduced in 2010 an animated CAPTCHA, called \textbf{HelloCAPTCHA}, freely available through the developers' website. 
In general, the HelloCAPTCHA challenge consists of a sequence of six characters presented in an animated GIF image. In some challenges, the characters change position and orientation, and in others, they are not all visible at the same time. The idea behind such a scheme is to spread the information over multiple animation frames to prevent a typical OCR attack over a single frame.
\textbf{NuCaptcha} is another animated CAPTCHA scheme \cite{ref37}. The challenge consists of a video with scrolling text in white font, followed by three random red characters moving across a dynamic background. The user is required to type the moving red characters to solve the CAPTCHA.
\textbf{Dracon CAPTCHAs} \cite{ref6} are animated visual Flash CAPTCHAs. The challenge consists of recognizing five characters displayed at fixed locations and randomly altered by using fade and blur effects. The animation is enriched with noise, e.g., random falling bars in the foreground or small text characters in the background.
\textbf{KillBot Professional} version \cite{ref33} is a commercial animated CAPTCHA that claimed among its client the United States Federal Government. In detail, the users have to recognize five moving characters displayed in a noisy foreground and background that are composed of lighter colors than the main text characters.
\textbf{Atlantis CAPTCHA} \cite{ref33} is an animated CAPTCHA used on the Atlantis website\footnote{Atlantis-caps.com}. In such a CAPTCHA, users need to recognize six moving characters among others that are continuously changing their color.

\subsection{Image-based CAPTCHAs}\label{sec:img captcha}
An alternative to text-based CAPTCHA schemes
are image-based ones. In these schemes, the challenge presented to the user is generally based on understanding a written text describing a task that needs an additional image classification or recognition task to be completed. The textual part has language dependencies. 
The user interaction or the gesture required to solve the challenge may differ from a scheme to another, therefore, we suggested a classification based on those differences, identifying six different types, as shown in Table ~\ref{table:imageCAPTCHA} and described in the following.

%%%%%%%%%%%%%%%%%%%%%%% TABLE 2 GOES HERE %%%%%%%%%%%%%%%%%%%%%%

\input{tables/table2}

\subsubsection{Click-based CAPTCHAs}\label{sec:click captcha}
%This type of CAPTCHA displays an image along with an instruction for the users to click on the mentioned position or areas.

%A typical example is \textbf{Implicit CAPTCHA} \cite{ref60}, where users are required to click on a specific static place on an image according to a given instruction, for instance "Click on the climber's glasses." or "Click on the logo on the climber's arm.". The major problem with such CAPTCHA is that the challenge cannot be generated automatically and thus it requires human effort to design it. Recently, a new image-based CAPTCHA, called \textbf{SACaptcha}, has been introduced by Tang et al.\cite{ref61}. Users are required to click on some regions in the image that have a specific shape mentioned in the challenge description to pass the CAPTCHA test.

This type of scheme shows an image and a text that explains where the user needs to click to complete the challenge.
A typical example is \textbf{Implicit CAPTCHA} \cite{ref60}, where the users are required to click on a specific static place on an image according to the given instruction, e.g., ``Click on the climber's glasses'' or ``Click on the logo on the climber's arm''. 

The major limitation of such a CAPTCHA scheme is that the challenge cannot be generated automatically, and thus it requires the human intervention to generate a new instance. 
Recently, a new image-based CAPTCHA, called \textbf{SACaptcha}, has been introduced by Tang et al.\cite{ref61}. Users are required to click on some regions in the image that have a specific shape mentioned in the challenge description to pass the CAPTCHA test.
 
\subsubsection{Sliding image-based CAPTCHAs}\label{sec:slider captcha}

In sliding image-based CAPTCHAs, users are required to use the slider to solve an image-based challenge such as adjusting the orientation of an image, selecting the correct form of an image, or moving a fragment of an image to the correct location.

For instance, \textbf{WHAT's Up CAPTCHA} \cite{ref50} presents three randomly rotated images to the users and asks them to use the slider to rotate the images to their upright position. The success rate of a random guess depends on the tolerance of accepted answers. 
According to the data reported in \cite{ref50}, the success rate of a random guess on one image is 4.48\%, but it decreases to 0.009\% for three images. 
Slide-to-fit CAPTCHA \cite{ref5} by \textbf{Minteye} presents a distorted image through a swirl filter with a small slider below the image. Users have to move the slider until the user sees the undistorted version of the image.  \textbf{Tencent CAPTCHA} asks the users to drag the slider until two puzzle pieces match. One of these puzzle pieces represents the target region in the image, where the users have to place the other piece of the puzzle to have a complete image.

\subsubsection{Drag and drop based CAPTCHAs}\label{sec:dragdrop captcha}

The Drag and drop CAPTCHA scheme requires the users to combine or reorder image pieces by dragging and dropping them to form a complete picture. 

For instance, \textbf{Garb CAPTCHA} \cite{ref55} presents an image divided into four pieces randomly shuffled. To pass the CAPTCHA test, users have to reorder them to reconstruct the original image. 
Similarly, \textbf{Hamid Ali et al.} \cite{ref57} introduced a puzzle-based CAPTCHA. The challenge consists of dragging and dropping four images or pieces of the same image into an empty grid of four cells. To pass the CAPTCHA test, the position of each image in the grid should be the same as in the reference image. 
\textbf{Gao et al.} \cite{ref56} proposed an image-based CAPTCHA that uses the Jigsaw puzzle. Their CAPTCHA displays an image divided into pieces (i.e., 9, 16, or 25 depending on security level), but only two are not in the original positions. Users have to identify the two pieces and drag one over the other to swap them to solve the puzzle. 
\textbf{Capy CAPTCHA} \cite{ref53} asks the users to drag one puzzle piece into the correct location within the challenge image. 
The puzzle void is filled with a fraction from the same or another image rather than a random color. 
\textbf{KeyCAPTCHA} \cite{ref54} shows an incomplete image along with three puzzle pieces and asks the users to assemble the image as they see it in the reference image displayed in the upper right corner of the frame. The reference image is shown with a small resolution, and it disappears once the cursor is inside the frame. To pass the CAPTCHA test, the users have to drag and drop the three puzzle pieces in their correct position.

\subsubsection{Selection-based CAPTCHAs}\label{sec:text captcha}

Selection-based CAPTCHA schemes ask users to select candidate images from sets of images. The task can be described with text only or with text and a sample image.

A typical CAPTCHA of this kind is \textbf{Asirra} \cite{ref38}, which displays 12 images of cats and dogs and asks users to select all cat images among them. 
Similarly, \textbf{HumanAuth CAPTCHA} \cite{ref70} asks the users to select all images with natural content. 
It is based on humans' ability to distinguish between images with natural content (e.g., tree, river) and artificial one (e.g., car, watch). 
In contrast to Asirra and HumanAuth CAPTCHA, \textbf{SEMAGE} (SEmantically MAtching imaGEs) CAPTCHA \cite{ref39} asks users to select semantically related images from a given image set. Thus, the user is required to recognize the content of each image and then understand and identify the semantic relationship between a subset of them. 

In 2014, Google introduced the second version of reCAPTCHA based on behavior analysis, called \textbf{ ``No captcha reCAPTCHA''} \cite{ref71}, \cite{ref72}. In this version the system analyzes the browser environment (e.g., browser history, cookies, etc.) and evaluate the risk of being confronted with a bot; if the risk is considered high, then the page displays a selection-based CAPTCHA, otherwise checking a checkbox is enough. 
The selection-based CAPTCHA challenge consists of a sample image with a keyword describing the content of the image and 9 candidate images. The user is required to select images that are similar to the sample to pass the challenge. 

\textbf{ Facebook's image CAPTCHA} follows the same approach of reCAPTCHA except for the sample image. 
To pass the challenge, users have to select the images that correspond to the description (i.e., hint) from twelve images with different content. Afterward, Google introduced other variations of image-based reCAPTCHA that ask the user to select images with vehicles, houses, street signs, or other specific objects.  

Among others, several selection-based CAPTCHAs rely on face images for their challenges. 
For instance, \textbf{Avatar CAPTCHA} \cite{ref43} requires users to choose avatar faces from a set of 12 grayscale images composed of a mix of human and avatar faces. Other face-based image CAPTCHAs are \textbf{FR-CAPTCHA} \cite{ref46} and \textbf{FaceDCAPTCHA} \cite{ref45}.
FR-CAPTCHA asks users to select two face images of the same person displayed in a complex background. Differently, FaceDCAPTCHA requires users to identify the visually distorted real human faces among nonhuman face images. Unlike Avatar, the human face images used in FR-CAPTCHA and FaceDCAPTCHA are rotated, distorted, or embedded in a complex background.

\subsubsection{Drawing-based CAPTCHAs}\label{sec:draw captcha}

The CAPTCHAs schemes belonging to this category distinguish computers and human beings thanks to a drawing challenge. 

Shirali-Shahreza has introduced the first drawing-based CAPTCHA, named \textbf{Drawing CAPTCHA} in 2006 \cite{ref62}. Users are required to draw lines to connect diamond-shaped dots. 
These dots are displayed on a screen with noisy background, so users have to identify them first. 
Another CAPTCHA that falls into this category is \textbf{VAPTCHA} (Variation Analysis-Based Public Turing Test to Tell Computers and Humans Apart)\cite{ref68}. The VAPTCHA challenge consists of an image containing a randomly generated reference trajectory. Users are required to draw a resemblant trajectory to match the reference trajectory to complete the verification. If the matching degree is equal to or higher than the minimal match degree defined by the system, users are classified as humans, otherwise they are assumed to be bots. Similarly, \textbf{MotionCAPTCHA} \cite{ref69} asks users to draw a shape similar to the one displayed in the challenge box.

\subsubsection{Interactive-based CAPTCHA}\label{sec:mov cursor captcha}

CAPTCHA schemes in this category rely on the user's interaction through mouse movement or swiping gesture to discover a secret position in an image. This position represents the answer to the challenge and it is revealed only after the user's interaction. 

For instance, Conti et al. \cite{ref59} proposed a new CAPTCHA scheme, called \textbf{CAPTCHaStar}. The proposed CAPTCHA leverages the human ability to recognize shapes in a confusing environment. The underlying assumption is that a machine cannot easily emulate this ability. The CAPTCHaStar challenge consists of white pixels, called stars, randomly mixed during the generation of the challenge. The position of these stars changes according to the position of the cursor. To pass the CAPTCHA test, users have to move the cursor until the stars aggregate in a recognizable shape, then, click on the left mouse button to send the cursor coordinates to the server. If the cursor is close to the secret position, users are considered as humans. 
On mobile devices, CAPTCHaStar requires swiping the fingers to move the cursor and tapping the ``check'' link to submit the final answer. 

Similarly, Okada et al. \cite{ref65} introduced \textbf{Noise CAPTCHA}, which is composed of two noisy images with different sizes and a hidden object or message in a specific position in the image. To pass the CAPTCHA test, users have to move the small noisy image over the large image until the hidden object appears, then click on the ``submit'' button. Similar to CAPTCHaStar, users are considered as humans when they identify the correct (secret) position at which the object or the image becomes visible. 

Thomas et al. \cite{ref66} propose \textbf{Cursor CAPTCHA}, which displays five cursor images in a randomly generated image and customizes the cursor image of the mouse pointer. Then, the CAPTCHA asks users to overlap the mouse pointer on an identical cursor image to pass the challenge.  At the beginning of the test, users see six cursor images in which two of them are identical, but they are unable to identify the target position until they move the mouse. 

\subsection{Audio-based CAPTCHAs}\label{sec:audio captcha}
Audio-based CAPTCHA schemes were initially proposed as an alternative to visual CAPTCHAs for people who have a visual impairment. To pass the test, they are required to type what they have heard.  

One of the most popular audio-based CAPTCHA was the \textbf{audio reCAPTCHA} proposed by researchers at Carnegie Mellon University and later acquired by Google. 
To pass the CAPTCHA challenge, users have to recognize eight spoken digits with a background noise composed of human voices speaking backward at varying volumes. Audio reCAPTCHA accepts only one mistake in one of the digits to solve the challenge. 

Nevertheless, Sauer et al. \cite{ref78} showed that this CAPTCHA scheme represents a hard task for blind users. Indeed, their usability study involving six blind participants shows that the participants were able to complete only 46\% of the tasks correctly. 

Similarly, many popular websites implement audio CAPTCHAs that rely on listening to a random sequence of digits. 
For instance, \textbf{e-Bay Audio CAPTCHA} consists of six digits spoken in different voices with regular background noise. \textbf{Microsoft CAPTCHAs} are composed of ten digits spoken in different voices with regular background noise consisting of several simultaneous conversations. \textbf{Yahoo CAPTCHA} asks the users to type seven digits that follow three beeps spoken by a child with background noise consisting of other children's voices. The \textbf{Audio reCAPTCHA} version used in 2013, asks the users to identify all digits presented in the challenge composed of three clusters. Each cluster contains three or four overlapping digits. In 2017, Google released a new version of \textbf{reCAPTCHA} with ten spoken digits and background noise. The available experiences of Audio-based CAPTCHAs are summarized in table \ref{table:videoCAPTCHA}.

\subsection{Video-based CAPTCHAs}\label{sec:Video captcha}

CAPTCHA schemes in this category reproduce a short video and then propose a textually described challenge that requires some level of comprehension of the video content.

For instance, \textbf{Kluever et al.} \cite{ref85} proposed a CAPTCHA that asks the user to watch a video and provide three words that best describe the video. Similarly, Shirali-Shahreza et al. proposed \textbf{Motion captcha} \cite{ref86} that asks the users to watch a video, then they have to select the sentence that describes the motion of the person in the video.

The most common implementations of Video-based CAPTCHAs are reported in table \ref{table:videoCAPTCHA}.

%\subsubsection{Video as an answer}\label{sec:video as answer}
%CAPTCHAs in this category display the challenge in a form of text or image and asks the users to record a video while they are performing the requested task.WEBCAM CAPTCHA \cite{ref102} for instance asks the users to record a video with the device camera while performing a gesture using their face, hand or other body part. Similarly, FATCHA \cite{ref103} asks the users to record a video while performing a gesture such as showing the left ear or shaking the head. FATCHA can be used as well as liveness test for face authentication systems.

%%TODO remove newpage
%\newpage
%\onecolumn
%\afterpage{
\begin{table*}[h!]
\footnotesize
\centering

% with vertical tables
%\begin{longtable}[]{|m{0.13\textwidth}|m{0.11\textwidth}|m{0.16\textwidth}|m{0.18\textwidth}|m{0.03\textwidth}|m{0.22\textwidth}|}
%\twocolumn
%without vertical lines
\begin{tabular}[]{m{0.13\textwidth}m{0.20\textwidth}m{0.20\textwidth}m{0.03\textwidth}m{0.28\textwidth}}

\hline  \textbf{Captcha Type} & \textbf{Captcha Scheme} & \textbf{Sample} & \textbf{Year} & \textbf{Challenge Description} \\ \toprule

%\endhead

%%%%%%%%%%%%%%%%%%%%%%%%%%%%%%%%%%%%%%%%%%%%%%%%%%

% Video-based

\multirow{4}{*}[-18pt]{Video-based}  & Kluever el al \cite{ref85} & \includegraphics[width=3cm, height=2cm]{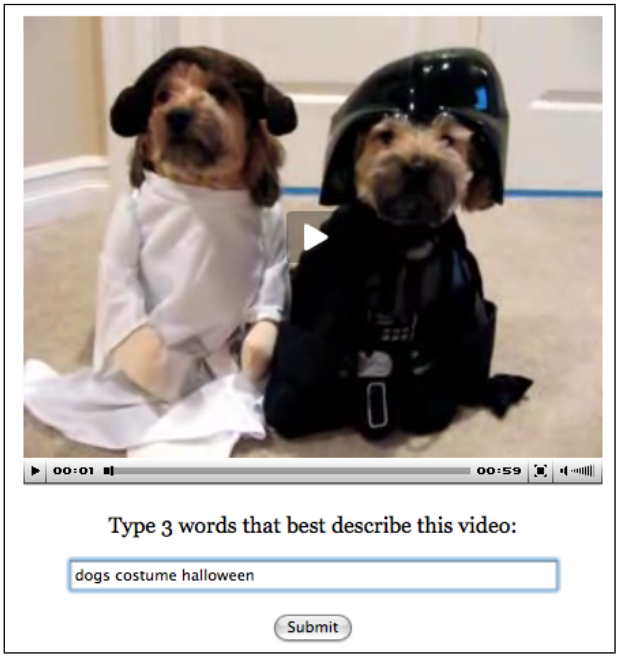} & 2009 & Watch a video and provide three words that best describe the video \tabularnewline
%\cline{3-6}
\cmidrule{2-5}

 &   Motion captcha \cite{ref86}  &  \includegraphics[width=3cm]{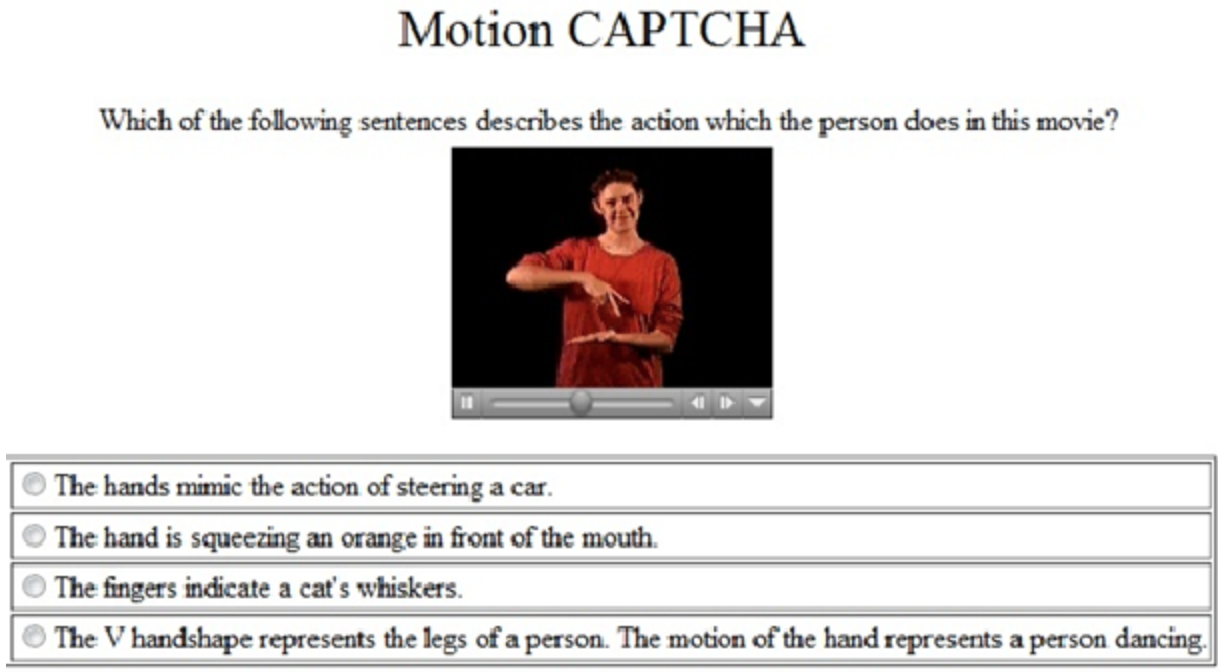}  & 2008 & Select the sentence that describes the motion of the person in the video\\ 
 
\midrule
%\hline

\multirow{6}{*}[-68pt]{Audio-based}  & Audio ReCAPTCHA (non-continous) &  &  2008 & Recognize eight spoken digits with background noise  consisting of human voices speaking backwards at varying volumes \\ 
\cmidrule{2-5}

  & e-Bay audio CAPTCHA  &  &  & Recognize six digits spoken in different voices with regular background noise    \\ \cmidrule{2-5}
 
  & Microsoft CAPTCHA    &  &  & Recognize ten digits spoken in different voices with regular background noise consisting of several simultaneous conversations     \\ 
 \cmidrule{2-5}
 
  & Yahoo CAPTCHA        &  &  & Recognize seven digits that follow three beeps spoken by a child with background noise consisting of other children's voices       \\ 
 \cmidrule{2-5}
 
  & Audio reCAPTCHA (Continuous)    &  & 2013 & Identify all digits presented in the challenge that consist of three clusters and each cluster contains three or four overlapping digits \\ 
 \cmidrule{2-5}
 
  & Audio ReCAPTCHA (version 2017)  &  & 2017 & Recognize ten spoken digits with background noise \\

\bottomrule

\end{tabular}
\caption{A taxonomy of video and audio-based CAPTCHAs }
\label{table:videoCAPTCHA}

\end{table*}

% \begin{tabular}{|m{0.13\textwidth}|m{0.12\textwidth}|m{0.13\textwidth}|m{0.16\textwidth}|m{0.05\textwidth}|m{0.26\textwidth}|} 
%\end{tabular}
%\twocolumn
%\clearpage

\subsection{Math-based CAPTCHAs}\label{sec:math captcha}
%CAPTCHAs in this category ask the users to solve a challenge based on mathematical problem.

%A typical example of Math-based CAPTCHA is \textbf{Arithmetic CAPTCHA} which rely on basic arithmetic operations such as (+,*,-). To solve the challenge users have to enter the results of a simple math operation such as "2+1= "to prove that they are human. 

%Similar to text-based CAPTCHA, using regular characters makes this CAPTCHA easy to break which leads some developers to design arithmetic CAPTCHA based on segmentation resistance mechanism. Unlike Arithmetic CAPTCHA, \textbf{QRBGS CAPTCHA} \cite{ref109} usually asks the users to solve complex equation that involve trigonometric and differential functions. The main problems with such CAPTCHA is that it suppose that all users have advanced knowledge in mathematics and it requires long time for solving the challenge.

CAPTCHA schemes in this category ask the users to solve a challenge based on a mathematical problem.

A typical example of Math-based CAPTCHA is \textbf{Arithmetic CAPTCHA} that relies on basic arithmetic operations such as (+,*,-). To solve the challenge, users have to enter the results of a simple math operation such as ``2+1= '' to prove that they are human. 
Unlike Arithmetic CAPTCHA, \textbf{QRBGS CAPTCHA} \cite{ref109} usually asks the users to solve a complex equation that involves trigonometric and differential functions. The main problem with such kind of CAPTCHAs is that it assumes that all users have advanced knowledge in mathematics, and it requires a long time to solve the challenge.

\subsection{Slider CAPTCHAs}\label{sec:slider captcha}

Slider CAPTCHA is another type of CAPTCHA scheme that relies only on the sliding gesture. Unlike sliding image-based CAPTCHAs previously described, image recognition is not part of the challenge. Users have only to move the slider across the screen to prove they are human. 

For instance, the CAPTCHA used by \textbf{Taobao.com}, which is a Chinese online shopping website owned by Alibaba, asks the users to drag the slider from the start to the end of the sliding bar to verify whether they are human or not. Similarly, CAPTCHA used by \textbf{TheyMakeApps.com} asks the users to move the slider to the end of the line to submit a form \cite{ref104}. This type of CAPTCHA has been widely adopted due to its ease of use.

Some well known examples of Math and Slider-based CAPTCHAs are reported in table \ref{table:mathCAPTCHA}.
%\newpage\onecolumn
%\afterpage{
\begin{table*}[h!]
\footnotesize
\centering

% with vertical tables
%\begin{longtable}[]{|m{0.13\textwidth}|m{0.11\textwidth}|m{0.16\textwidth}|m{0.18\textwidth}|m{0.03\textwidth}|m{0.22\textwidth}|}
%\twocolumn
%without vertical lines
\begin{tabular}[]{m{0.10\textwidth}m{0.23\textwidth}m{0.20\textwidth}m{0.03\textwidth}m{0.30\textwidth}}

\hline  \textbf{Type} & \textbf{Scheme} & \textbf{Sample} & \textbf{Year} & \textbf{Challenge Description} \\ \toprule

%\endhead

%math-based captcha
%\hline
\multirow{3}{*}[-5pt]{Math-based} &  Arithmetic CAPTCHA &  \includegraphics[width=3cm]{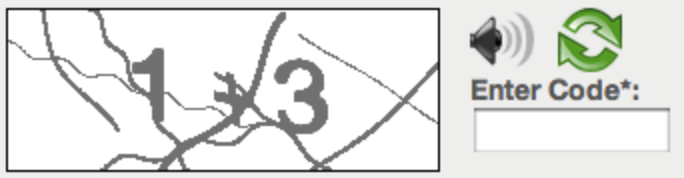} &  & Enter the result of the math operation \\ 
\cline{2-5}

&  QRBGS CAPTCHA \cite{ref109} & \shortstack[l]{\\ \includegraphics[width=3cm]{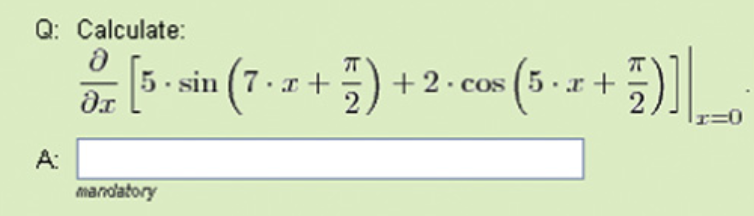}}  & 2008 & Enter the result of a complex mathematical equation \\
\midrule
%%%%%%%%%%%%%%%%%%%%%%%%%%%%%%%%%%%%%%%%%%%%%%%%%%%

%Slider-based captcha

\multirow{3}{*}[-8pt]{Slider-based} &  Taobao &  \includegraphics[width=3cm]{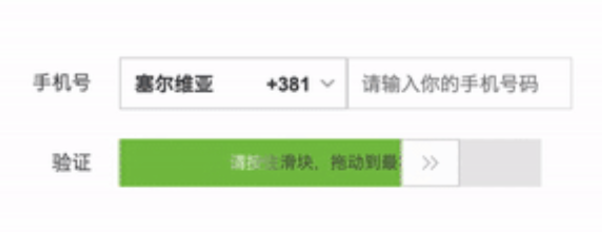}  &  & Drag a slider from the start to the end of the sliding bar \\
\cline{2-5}

&  TheyMakeApps CAPTCHA \cite{ref104} & \shortstack[l]{\\ \includegraphics[width=3cm]{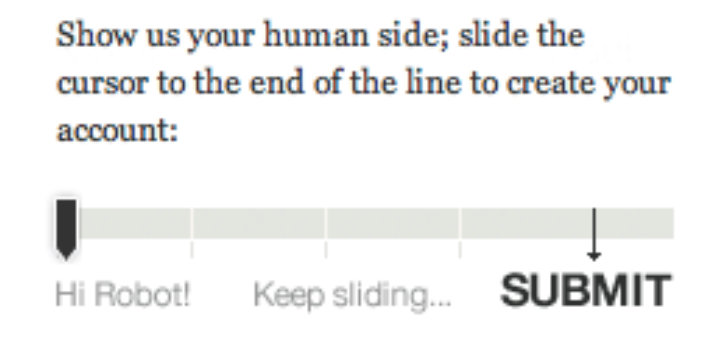}}  & 2010 & Move the slider to the end of the line \\

\bottomrule

\end{tabular}
\caption{A taxonomy of Math and slider-based CAPTCHAs }
\label{table:mathCAPTCHA}

\end{table*}
%\twocolumn
%\clearpage

\subsection{Game-based CAPTCHAs}\label{sec:game captcha}

Game-based CAPTCHA schemes have emerged as an alternative that tries to make the task of solving CAPTCHAs a fun activity for the users. These CAPTCHAs are based on the assumption that humans -- unlike automated systems -- can understand the rules of a game and solve the challenge.
Users are required to solve a straightforward game that is often based on image semantics. There are also attempts to make the users enjoy solving math-based CAPTCHAs by offering games such as tic-tac-toe and a dynamic roll-dice game.

A well-known game-based CAPTCHA is \textbf{PlayThru CAPTCHA} \cite{ref81} designed by a startup called ``Are you a human''. 
The challenge requires moving some dynamic objects that have a semantic connection with the static target image. For instance, users might be asked to place food in the refrigerator or feed a baby. 
Mohamed et al. \cite{ref80} developed four \textbf{Dynamic Cognitive Games (DCG)} similar to PlayThru, in order to investigate both its security and usability. 
Depending on the game, users are required to drag and drop dynamic objects to match them with others (e.g., match objects with similar shapes) or place them in specific regions (e.g., place the ships on the sea). 
Their usability study shows that all the four games last less than 10 seconds, and all the participants successfully completed the games within the time out. 
Regarding the error rate per drag and drop, the authors noticed that the visual matching tasks are less error-prone than the semantic matching tasks. 

Another example of game-based CAPTCHA  is \textbf{SweetCAPTCHA}. Also in this case, the users are required to drag and drop an image with a semantic connection with the target image. For example, users need to drag milk to a cup of coffee, drag the player to the guitar, or drag chopsticks to sushi.
Another example is \textbf{Tic Tac Toe CAPTCHA} that proposes to the user an almost complete the tic-tac-toe game, where users need a single move to win the game and get 3 Xs in a row.

Some CAPTCHA designers have tried to have users having fun when they solve CAPTCHAs based on a mathematical problem. A typical example is \textbf{Dice CAPTCHA} (i.e., Homo-sapiens Dice version) \cite{ref82}, where users are required to roll some dice and then compute the sum of the digits appearing on them. If the entered sum is correct, the users are considered humans. 

A detailed taxonomy of the most common game-based CAPTCHAs is reported in table \ref{table:gameCAPTCHA}.

%\onecolumn
%\afterpage{
\begin{table*}[h!]
\footnotesize
\centering

% with vertical tables
%\begin{longtable}[]{|m{0.13\textwidth}|m{0.11\textwidth}|m{0.16\textwidth}|m{0.18\textwidth}|m{0.03\textwidth}|m{0.22\textwidth}|}
%\twocolumn
%without vertical lines
%\begin{longtable}[]{m{0.13\textwidth}m{0.16\textwidth}m{0.18\textwidth}m{0.03\textwidth}m{0.22\textwidth}}
\begin{tabular}[]{m{0.22\textwidth}m{0.20\textwidth}m{0.03\textwidth}m{0.40\textwidth}}
%\hline \textbf{CAPTCHA Type} & \textbf{Captcha Scheme} & \textbf{Sample} & \textbf{Year} & \textbf{Challenge Description} \\ 
\hline  \textbf{Scheme} & \textbf{Sample} & \textbf{Year} & \textbf{Challenge Description} \\ \toprule

%\endhead

%%%%%%%%%%%%%%%%%%%%%%%%%%%%%%%%%%%%%%%%%%%%%%%%%%%

% GAme-based

%\multirow{6}{*}[-70pt]{Game-based} & PlayThru \cite{ref81} &  \includegraphics[width=3cm]{otherimages/Playthru.png}  & 2012 & Play a simple game that consist of moving specific dynamic objects to a specific place according to the image semantics\\ 
PlayThru \cite{ref81} &  \includegraphics[width=3cm]{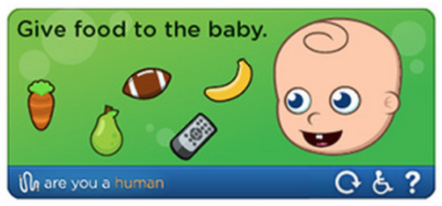}  & 2012 & Play a simple game that consist of moving specific dynamic objects to a specific place according to the image semantics\\ 
%\cline{3-6}
\cmidrule{1-4}

   DCG CAPTCHAs \cite{ref80} &  \includegraphics[width=3cm]{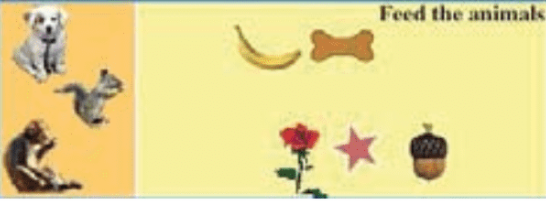}  & 2014 & Depending on the challenge description, drag and drop objects to match them with others or place them in specific regions\\ 
%\cline{2-6}
\cmidrule{1-4}
%\midrule
  SweetCAPTCHA &  \includegraphics[width=3cm]{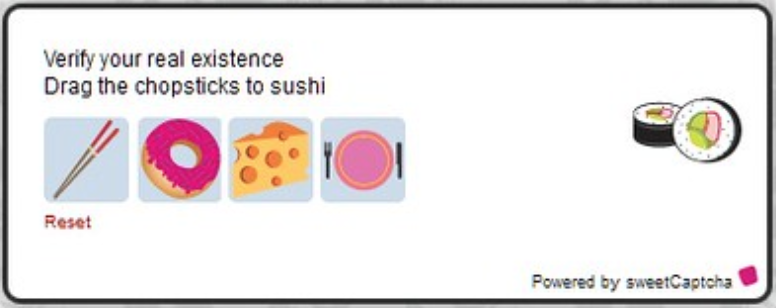}  & 2011 & Drag specific static images to match them with the target image\\ 
\cmidrule{1-4}

   Dice CAPTCHA \cite{ref82} &  \includegraphics[width=3cm]{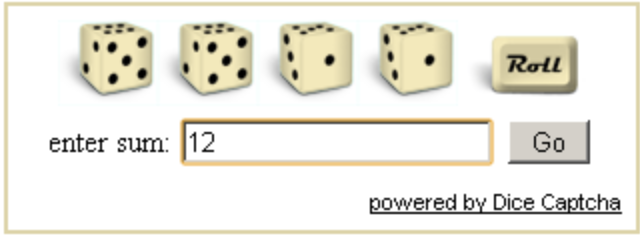}  & 2010 & Click on ``Roll'' to roll the dices, then enter the sum of  the numbers appearing on the dices\\ 
 \cmidrule{1-4}

  Tic tac toe CAPTCHA &  \includegraphics[width=3cm, height=3cm]{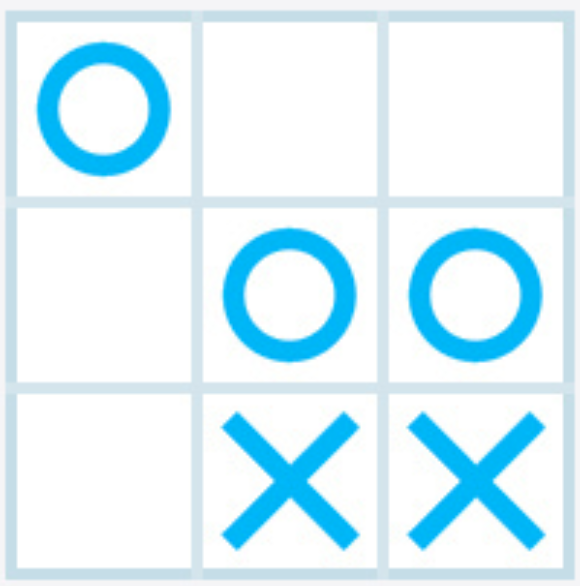}  & 2011 & Complete the game by tapping into the correct position to get a line of 3 Xs (or Os) \\ 
%\cline{3-6}

\bottomrule

\end{tabular}
\caption{A taxonomy of game-based CAPTCHAs }
\label{table:gameCAPTCHA}

\end{table*}

%\twocolumn
%\clearpage

%\subsubsection{Math-based game CAPTCHAs:}\label{sec:math game captcha}
%CAPTCHAs in this category try to make users having fun when they solve CAPTCHAs based on mathematical problem.

%\subsection{Biometric-based CAPTCHAs}\label{sec:puzzle captcha}

%This new direction of CAPTCHA uses one or multiple biometric modalities which belongs to physiological biometrics, behavioral biometrics or a combination between the two. 
%The most proposed biometric CAPTCHAs proposed in the literature focus on behavioral biometrics while some of them use physiological or combination between behavioral and physiological . 

%This new direction of CAPTCHA analyses the users behavior when they are performing a specific task which can be requested explicitly from the user such as asking the user to click on checkbox or to drag the slider to right or it can be implicit. This latter do not require from the users to solve a challenge instead it leverages the existing tasks such as user's click on the submit button required to send a form or a comment to distinguish between humans and bots. 

\subsection{Behavior-based CAPTCHAs}\label{sec:click behavior captcha}

CAPTCHA schemes in this category employ behavioral biometrics such as keystroke dynamics, mouse dynamics, swipe dynamics, and eye movement to distinguish between humans and bots. Most of the proposed schemes %The most proposals\footnote{to my knowledge the most proposals is not correct English, "most of the proposals" or, as we are going to change it "most of the proposed schemes"}
involve mouse/swipe dynamics with conventional CAPTCHA schemes (e.g., image-based or game-based).

As an example, Acien el al. \cite{ref106} proposed in 2020 \textbf{BeCAPTCHA-Mouse}, which asks the user to solve an image-based CAPTCHA similar to reCAPTCHA V2. However, such a scheme analyzes the mouse trajectories performed during the task to distinguish between humans and bots. 
Similarly, \textbf{Gametrics} \cite{ref105} asks the users to solve a Dynamic Cognitive Game CAPTCHA. During the drag and drop operations requested to solve the challenge, the CAPTCHA collects the mouse movement features to distinguish between human and automated systems. 

In addition, \textbf{GEETest} and \textbf{Netease} \cite{ref42} ask the users to solve a sliding image-based CAPTCHA similar to Tencent CAPTCHA. In detail, the users need to complete an image by dragging the slider to match two puzzle pieces (one reflecting the missing part of the image, the other the correct position in the image).
Unlike Tencent CAPTCHA, users are considered humans only when both the puzzle pieces match and the sliding behavior is not considered suspicious. 

Furthermore, the same authors of BeCAPTCHA-Mouse proposed a variation for smartphones called \textbf{Be-CAPTCHA} \cite{ref84} that is based on a slider challenge. 
However, unlike traditional sliding tasks, the algorithm leverages swiping gestures and sensor data to detect human behavior.

Siripitakchai et al. \cite{ref88} proposed \textbf{EYE-CAPTCHA}, which asks the users to solve a math-based CAPTCHA relying on the eye movement. In detail, the challenge prompts a simple math operation in the center on the screen, along with four potential answers at the corners. To solve the challenge, the user has to locate the right answer and move it through his eyes to the center.

Unlike the above-mentioned behavioral CAPTCHAs, the \textbf{``No CAPTCHA reCAPTCHA''} (a.k.a., reCAPTCHA V2) deployed by Google in 2014 does not use a traditional CAPTCHA scheme to gather information on the user behavior. On the contrary it only requires to click on the ``I'm not a robot'' Checkbox. 
However, in the background, information related to user's behavior (e.g., the mouse movement, where the users click, how long they linger over a checkbox) along with other information such as the installed plugins, the language of the browser and cookies are collected and analyzed by an engine that evaluates the risk of being confronted with a bot. 
If the user is classified as human, no additional tasks are required. Otherwise, the system prompts a traditional image-based reCAPTCHA as a second security layer. 

In 2017, Google released another variation of reCAPTCHA V2, called \textbf{Invisible reCAPTCHA}. As its name suggests, the challenge is invisible to the user. The verification process is performed in the background, and it is invoked when the user clicks on an existing button on the web page or by a JavaScript API call. 
Similarly to the ``No CAPTCHA reCAPTCHA'' approach, Invisible reCAPTCHA requires to solve the traditional image-based reCAPTCHA if and only if the risk analysis engine cannot recognize a human behavior with a given level of confidence.

A detailed taxonomy of the most common behavior-based CAPTCHAs is reported in table \ref{table:behaviorCAPTCHA}.

%\onecolumn
%\afterpage{
\begin{table*}[h!]
\footnotesize
\centering

% with vertical tables
%\begin{longtable}[]{|m{0.13\textwidth}|m{0.11\textwidth}|m{0.16\textwidth}|m{0.18\textwidth}|m{0.03\textwidth}|m{0.22\textwidth}|}
%\twocolumn
%without vertical lines
\begin{tabular}[]{m{0.27\textwidth}m{0.21\textwidth}m{0.03\textwidth}m{0.40\textwidth}}

\hline  \textbf{Scheme} & \textbf{Sample} & \textbf{Year} & \textbf{Challenge Description} \\ \toprule

%\endhead

%% Behavior Based

    BeCAPTCHA-Mouse \cite{ref105} &  \includegraphics[width=3cm, height=3cm]{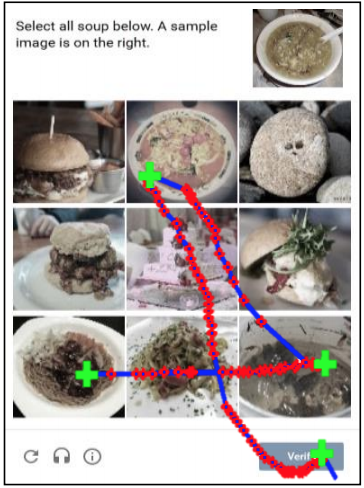} & 2020 & Solve a selection image-based CAPTCHA\\ 
%\cline{2-6}
\cmidrule{1-4}

    Gametrics \cite{ref105} &  \includegraphics[width=3cm, height=2cm]{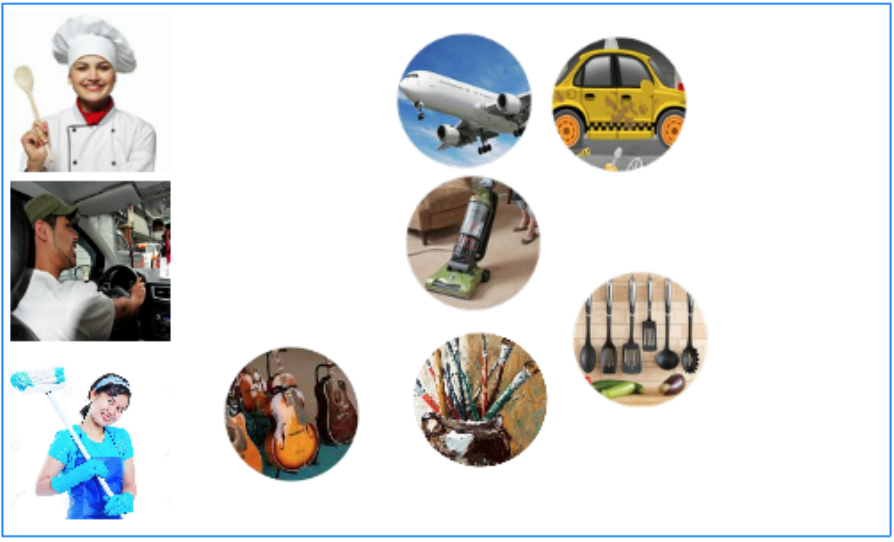} & 2016 & Drag-drop a subset of the moving objects to their orresponding targets which are static\\ 
%\cline{2-6}
\cmidrule{1-4}

   GEETest (geetest.com)&  \includegraphics[width=3cm, height=2cm]{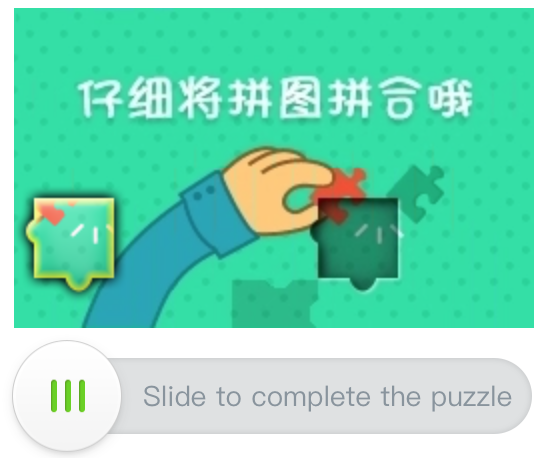}  & 2012 & Drag the slider until two puzzle pieces match \\ 
%\cline{3-6}
\cmidrule{1-4}

   Netease \cite{ref42} &  \includegraphics[width=3cm, height=2cm]{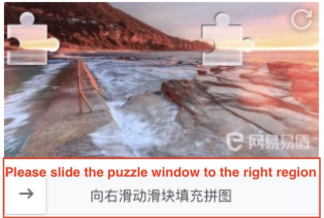}  &  & Drag the slider until two puzzle pieces match \\ 
%\hline
\cmidrule{1-4}

   Be-CAPTCHA \cite{ref84} &  \includegraphics[width=3cm]{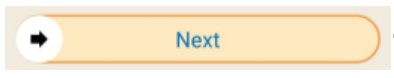}  & 2020 & Drag a slider from the start to the end of the sliding bar \\ 
%\cline{3-6}
\cmidrule{1-4}

  Eye-CAPTCHA \cite{ref88} & \includegraphics[width=3cm, height=2cm]{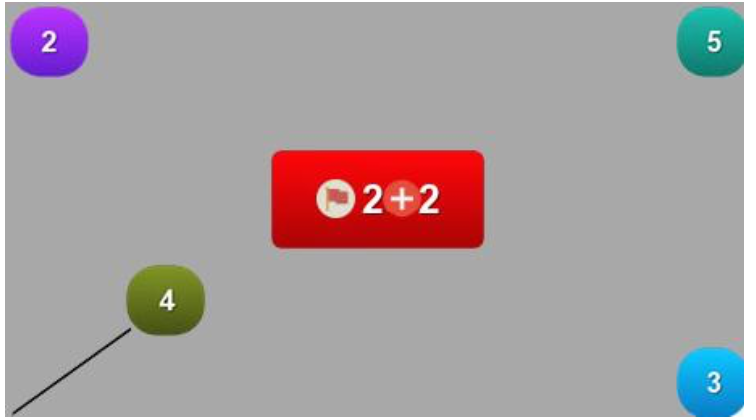} & 2017 & User locates the answer of a simple math operation displayed in the screen and move it using his eyes to the center\\
\cmidrule{1-4}

%\midrule
     No CAPTCHA reCAPTCHA \cite{ref72} &  \includegraphics[width=3cm]{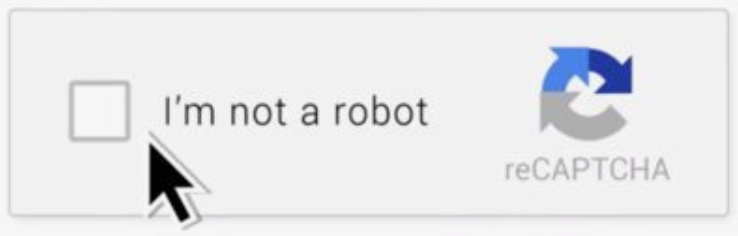}  & 2014 & Click on I'm not a robot Checkbox \\ 
%\cline{3-6}
\cmidrule{1-4}
    Invisible reCAPTCHA \cite{ref72} &  \includegraphics[width=3cm]{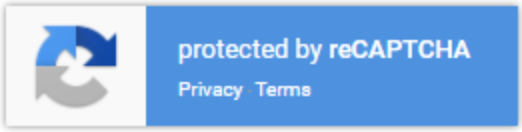}  & 2017 & No visible challenge, it is invoked via a Javascript API or  when the user clicks on an existing button on the website\\ 

\bottomrule

\end{tabular}
\caption{A taxonomy of behavior-based CAPTCHAs }
\label{table:behaviorCAPTCHA}
\end{table*}

\subsection{Sensor-based CAPTCHAs}\label{sec:sensor captcha}

%CAPTCHAs in this category usually are designed for mobile devices. Existing CAPTCHAs in the literature that leverage sensors can be divided into two categories, Physical CAPTCHAs which are mainly based on the sensors readings to distinguish between the human and bots and cognitive sensor-based CAPTCHA which are mainly based on cognitive tasks and use the sensors readings just as a way of input. 

The CAPTCHA schemes belonging to this category rely on the data gathered by one or more hardware sensors. These CAPTCHA schemes are typically designed for mobile devices that natively host sensors like gyroscope or accelerometer.
Sensors-based CAPTCHA schemes can be further divided into \emph{physical} and \emph{cognitive}. In the first case, the sensors' data are used to discriminate between a human and a bot. 
In the latter, the sensors only provide an input channel for the actions of the user.

A detailed taxonomy of the common available sensor-based CAPTCHA experiences is reported in table \ref{table:sensorCAPTCHA}.

\subsubsection{Physical CAPTCHAs}\label{sec:Physical captcha}
%Physical CAPTCHAs leverage sensors and human ability to perform a physical task to prove they are humans.

%To the best of our knowledge, the first physical CAPTCHA for mobile devices has been introduced by Guerar el al. \cite{ref96} in 2015. Their proposed CAPTCHA, called \textbf{CAPPCHA} (Completely Automated Public Physical test to tell Computers and Humans Apart) or \textbf{TiltToGo}, asks the users to tilt the device to a specific degree to prove they are humans. The idea behind this is that the bot as a piece of code can not perform a physical task such as moving the device. Using secure sensors, there is no need to randomize the challenge or to ask the users to perform a complex gesture. Therefore, authors suggested a simple gesture such as tilting the device to a specific degree which can be detected easily through motion sensors such as the accelerometer and gyroscope.

The first physical CAPTCHA for mobile devices has been introduced by Guerar et al. \cite{ref96} in 2015. The proposed CAPTCHA scheme, called \textbf{CAPPCHA} (Completely Automated Public Physical test to tell Computers and Humans Apart), requires the users to tilt the device to a specific degree to prove they are humans. 
The challenge exploits the impossibility for a software bot to perform a physical task such as moving the device. Furthermore, thanks to the use of dedicated hardware sensors, the CAPTCHA scheme does not require randomizing the challenge or executing sophisticated gestures.
Therefore, the authors suggested a simple gesture such as tilting the device to a specific degree, which can be detected easily through motion sensors such as the accelerometer and gyroscope.

%Similarly, in 2016, Hupperich el al. \cite{ref92} proposed \textbf{Sensor CAPTCHA} which asks the users to move the device to prove they are humans. Unlike CAPPCHA, Sensor CAPTCHA asks the users to perform a complex and awkward gesture such as hammering, fishing, drinking or turning the body while holding the mobile device. A limited number of gestures makes this kind of CAPTCHA susceptible to replay attack while, implementing more varieties of usable gesture is challenging. 

Similarly, in 2016, Hupperich el al. \cite{ref92} proposed \textbf{Sensor CAPTCHA} that asks the users to move the device to prove they are humans. Unlike CAPPCHA, Sensor CAPTCHA asks the users to perform a complex gesture such as hammering, fishing, drinking, or turning the body while holding the mobile device. 

In \cite{ref98}, the authors suggested \textbf{Pedometric CAPTCHA} that requires walking at least five steps to be considered humans. The idea behind this is to create an acceleration in the mobile device while the user is walking that cannot be generated by a bot. 
\textbf{Mantri et al.}\cite{ref99} proposed a CAPTCHA scheme that asks the users to move the device according to a specific pattern displayed on the screen. For instance, the user is required to write an ``S'' letter while holding the device and then press the ``submit'' button. 
Similarly, \textbf{Frank et al.} \cite{ref100} asks the users to move the device to perform a gesture that can be detected by the gyroscope, such as tilting the device, rotating the device or drawing a three-dimensional shape or letter while holding the device. 

In \cite{ref97}, Guerar et al. proposed \textbf{Invisible CAPPCHA} based on the same idea of CAPPCHA, although - as the name suggests - the challenge is invisible to the users. 
The authors noticed that most of the online services that require protection against automation abuses in mobile devices require the interaction with the touchscreen (e.g.,  fill a form, write a comment, tap on a button, perform the login). Such physical interactions cause micro-movements of the device that can be tracked by motion sensors such as the accelerometer. 
Based on their observation, they leveraged the implicit user's taps to make the challenge transparent to the users and thus more user-friendly. Unlike the Invisible reCAPTCHA designed by Google, Invisible CAPPCHA is based on humans' ability to perform a physical task and not on the way they perform the task. Also, the tap gesture is detected through sensors readings rather than touchscreen events that can be easily simulated by the bots \cite{ref137}. Furthermore, no sensitive data are provided to the server side as the interpretation of the sensor data is completely performed inside trusted hardware in the client side and thus it preserves the user's privacy.

\subsubsection{Cognitive sensor-based CAPTCHAs}\label{sec:cognitive captcha}

Similar to the traditional CAPTCHAs, Cognitive sensor-based CAPTCHAs ask the users to solve a cognitive challenge (e.g., recognizing an image, or solving a game, selecting images based on expert medical knowledge  \cite{ref161}), yet they use sensors as their input to solve the challenge rather than the conventional taping or swiping gestures. 
%Users are considered humans when they successfully solve the challenge. 
To this aim, we classified these CAPTCHAs as sensor-based CAPTCHA rather than including them in one of the categories mentioned above to highlight the current research trends.

A typical example of this category is \textbf{AccCAPTCHA} \cite{ref94}, where the challenge requires to play a simple game such as the rolling ball game. Thanks to the device's motion sensors, the user can move the ball to complete the game.

Yang et al \cite{ref79} proposed \textbf{GISCHA}, a game-based image semantic CAPTCHA for mobile devices. The challenge consists of a rolling ball and destination holes with different shapes. The direction of the rolling ball can be controlled by turning the mobile device to different angles. The users are considered as human if they successfully move the ball to the destination hole shaped like a circle. 
Similarly, the CAPTCHA designed by \textbf{Ababtain el al.} \cite{ref91} asks the users to solve a simple game to prove that they are humans, also in this case, using the sensors as their input. They suggested five games where all of them use one dynamic object and one or multiple static objects. To pass the test, the users have to move the dynamic object, so that it touches specific static objects which are considered as targets.

Recently, Feng el al. \cite{ref94} proposed \textbf{SenCAPTCHA} that is based on the difficulty of finding an animal facial key point. 
Such a CAPTCHA scheme proposes an image of an animal along with a small red ball. The users are required to tilt their devices to move the red ball into the center of that animal's eye. 
The idea behind using the sensor readings is to avoid the traditional input modalities (i.e., typing, selecting images) that can be inconvenient on devices with small screen sizes.

\subsection{CAPTCHAs for liveliness detection in authentication methods}\label{sec:auth captcha}

Today, one of the biggest problems that threats every website with a login is the use of malicious bots for credential stuffing and credential cracking. This is due to the availability of billions of breached credentials. Imperva \cite{ref136} reported that a recent credential stuffing attack lasted 60 hours and included 44 million login attempts. In the literature, there are many proposals that attempt to embed a form of CAPTCHA in the authentication methods to stop these attacks. 

In 2010, Stefan Popoveniuc \cite{ref101} proposed an authentication method called SpeakUP, for remote unsupervised voting. They added text-based CAPTCHA to voice biometrics. To log in, the voters are required to read out loud a 2D text CAPTCHA displayed on the screen that is associated with the candidate for whom they wish to vote. The voters are identified by the biometric characteristics of their voices. For further security, the author suggested to capture a video of the voter while solving the CAPTCHA. 

Recently, Uzun el al. \cite{ref89} proposed a Real Time CAPTCHA system called rtCaptcha for defending against automated attacks on facial authentication systems. Similar to SpeakUp CAPTCHA, once the authentication session start, users are required to take a video while pronouncing out loud the 2D text CAPTCHA presented as a challenge to prove they are humans. The session will time out if no response is received after a predefined period.

In \cite{ref135}, authors proposed BrightPass, an authentication method for mobile social media networks. They added liveliness detection mechanism to PIN/password in order to prevent the automated process of iterating through the entire password space and from testing all the stolen passwords. Their proposed mechanism leverages screen brightness, which 
cannot be captured by malicious programs, to tell the users when to input a correct PIN digit and when to input a misleading lie digit. 

In \cite{ref138}, \cite{ref139}, authors proposed a PIN-based authentication method for smartwatches that embeds a form of physical CAPTCHA. This mechanism uses the same principle behind CAPPCHA \cite{ref140}. Users have to physically rotate the bezel to a specific degree to input the PIN digits. Using a trusted hardware (i.e., the bezel) this mechanism prevents any automated program from performing a brute force or credential stuffing attacks. This mechanism can be also used separately from PIN-based authentication. Similarly, authors in \cite{ref141} leverage the rotation of the smartwatch digital crown to prevent automated attacks against the PIN code.

\newpage

%\onecolumn
\afterpage{%
\FloatBarrier
%\begin{table*}[h!]
\afterpage{
\footnotesize
\centering

% with vertical tables
%\begin{longtable}[]{|m{0.13\textwidth}|m{0.22\textwidth}|m{0.16\textwidth}|m{0.18\textwidth}|m{0.03\textwidth}|m{0.22\textwidth}|}
%\twocolumn
%without vertical lines
%original 13 e 20
\begin{longtable}[]{m{0.10\textwidth}m{0.25\textwidth}m{0.20\textwidth}m{0.03\textwidth}m{0.28\textwidth}}

\hline  \textbf{Type} & \textbf{Scheme} & \textbf{Sample} & \textbf{Year} & \textbf{Challenge Description} \\ \toprule

%\endhead

%%%%%%%%%%%%%%%%%%%%%%%%%%%%%%%%%%%%%

%Sensor-based

 \multirow{6}{*}[-85pt]{Physical} & CAPPCHA  \cite{ref96} & \includegraphics[width=2cm, height=3.5cm]{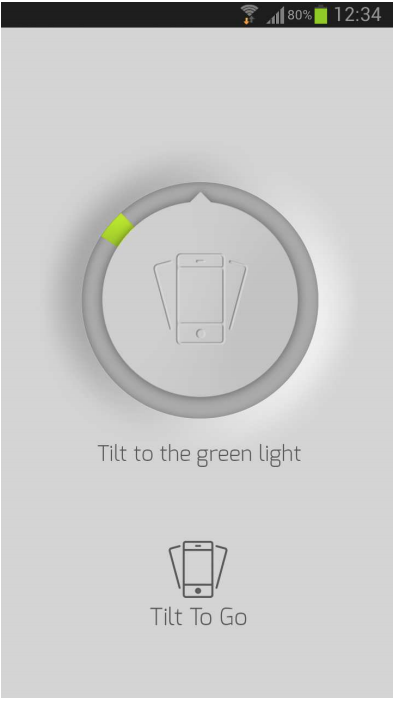} & 2015 & Tilt the device to a specific degree \\ 
\cmidrule{2-5} 

  & Pedometric CAPTCHA \cite{ref98} & \includegraphics[width=3cm, height=2cm]{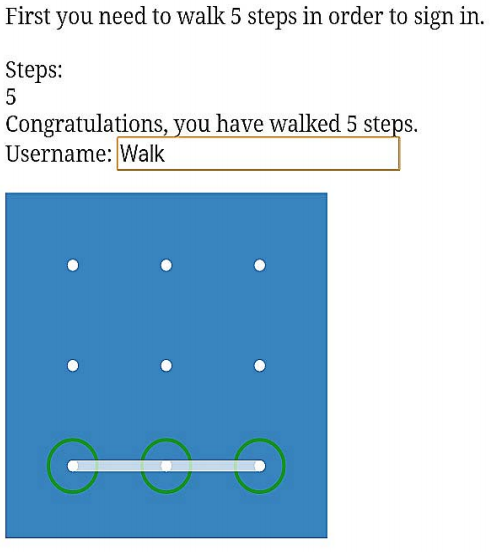} & 2017 & Walk at least 5 steps \\ 
  \cmidrule{2-5} 

 & Mantri et al. \cite{ref99} & \includegraphics[width=2cm, height=3cm]{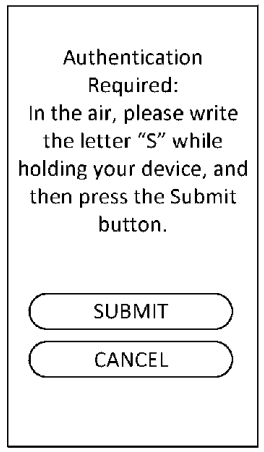}  & 2017 & Move the device according to a specified pattern displayed on the screen \\ 
\cmidrule{2-5} 

  & Sensor CAPTCHA \cite{ref92} &  & 2016 & Perform gestures such as hammering, fishing, turning the body while holding the mobile device \\ 
 \cmidrule{2-5}
 
  & Invisible CAPPCHA \cite{ref97} &  & 2018 & No task is required \\ 
 \cmidrule{2-5}
 
  & Frank et al. \cite{ref100} &        & 2018 & Move the device to perform an action (e.g., tilt the device, draw a shapes, letters or patterns) \\ 
 \cmidrule{1-5}
 
 \multirow{4}{*}[-55pt]{Cognitive } & AccCAPTCHA \cite{ref94} & \includegraphics[width=3cm, height=2cm]{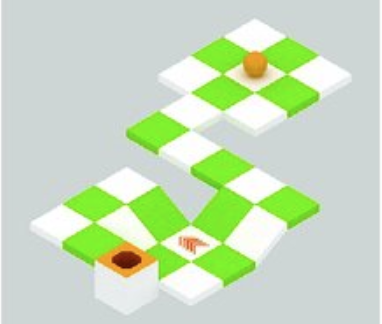} & 2013 & Play a simple rolling ball game or other well-known games (e.g., enigma, racing game) \\ 
 \cmidrule{2-5}
 
  & GISCHA \cite{ref79}  & \includegraphics[width=3cm, height=2cm]{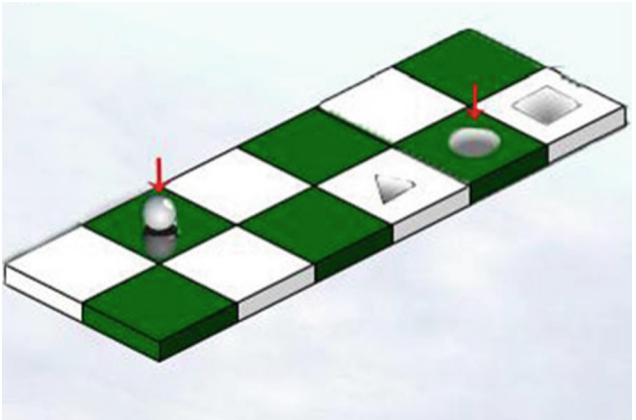} & 2013 & Play a simple game that consist of moving a ball to the destination hole with a specific shape \\ 
 \cmidrule{2-5}
 
  & SenCAPTCHA \cite{ref93} &  \includegraphics[width=3cm, height=3cm]{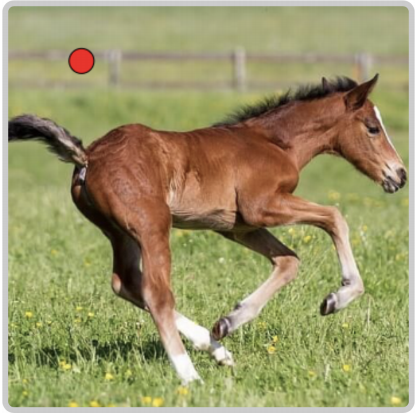} & 2020 & Identify the animal eye position, then tilt the device to move the ball to this position \\ 
 \cmidrule{2-5}
 
   &  Ababtain et al. \cite{ref91} &  \includegraphics[width=3cm, height=3cm]{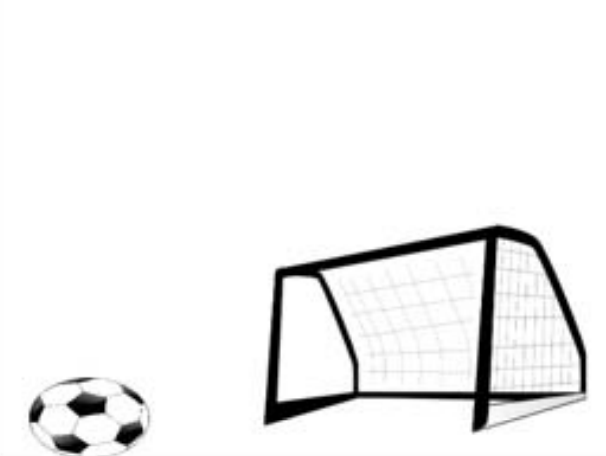}     & 2019 & Play a simple game that consist of one moving object (i.e., ball) and one or multiple target objects (e.g., Goal). Users move the device to match the ball with the target object \\
% \midrule
\bottomrule

\caption{A taxonomy of sensor-based CAPTCHAs }
\label{table:sensorCAPTCHA}
\end{longtable}
%\end{table*}
}
} %afterpage
\FloatBarrier
%\twocolumn
%\clearpage

% \begin{tabular}{|m{0.13\textwidth}|m{0.12\textwidth}|m{0.13\textwidth}|m{0.16\textwidth}|m{0.05\textwidth}|m{0.26\textwidth}|} 
%\end{tabular}

\section{Security of CAPTCHA schemes}\label{sec:attacks}

The different proposals of CAPTCHA schemes aim to discern between human and computing systems thanks to a challenge. Instead, from an attacker perspective, the goal is to break the CAPTCHA scheme, i.e., to solve the proposed challenge with an automated system and still being recognized as a human. 

The general process of breaking traditional CAPTCHAs can be divided into the following phases/stages: pre-processing, segmentation, and recognition. Pre-processing techniques (e.g., image binarization, image thinning, and noise removal) are usually used to remove background patterns, separate the foreground from the background, and eliminate noise before the segmentation and recognition phases \cite{ref144}. In some cases, extraction techniques are used before pre-processing \cite{ref34}, such as Pixel Delay Map (PDM), Catching Line (CL), and Frame Selection (FS). %\textcolor{red}{
Segmentation techniques are used to split the CAPTCHA image into segments that contain individual objects to facilitate recognition. Well-known techniques that have been used in breaking CAPTCHAs are vertical histogram, color-filling, snake segmentation \cite{ref144}, and JSEG.%} 

Many efforts have been put into breaking the different CAPTCHAs by the scientific community in the last years. To do so, attackers can rely on a set of attacking methodologies that can be grouped in:

\begin{itemize}
%    \item \textbf{Object Segmentation Attacks.} In this category, segmentation techniques are used to split the CAPTCHA image into segments that contain individual objects to facilitate recognition. Well-known techniques that have been used in breaking CAPTCHAs are vertical histogram, color-filling, snake segmentation \cite{ref144}, and JSEG. 
    
    \item \textbf{Object recognition attacks.} This type of attack includes object recognition attacks, pixel-count, dictionary and database attacks \cite{ref144}. The common techniques used for object recognition are pattern matching (e.g., shape context matching \cite{ref10}, correlation algorithm \cite{ref11}), OCR recognition, Scale-Invariant Feature Transform (SIFT) and, recently, deep learning. %\textcolor{red}{
    In particular, the most used deep learning models for CAPTCHA recognition are CNN, RNN, and LSTM-RNN \cite{ref20},\cite{ref150},\cite{ref151}.%}

    \item \textbf{Random Guess Attacks.} In this type of attack, attackers try to break the CAPTCHA scheme by guessing the correct answer. Therefore, CAPTCHAs with a small number of different challenges are vulnerable to this attack.
    
    \item \textbf{Human Solver Relay Attacks.} The bot forwards the CAPTCHA challenges to remote human workers to solve the CAPTCHAs in exchange for a small income. The human workers solve the challenges and send the correct responses to the bot that can solve the CAPTCHA accordingly.
\end{itemize}

\begin{figure*}[h!]
  \includegraphics[width=\textwidth]{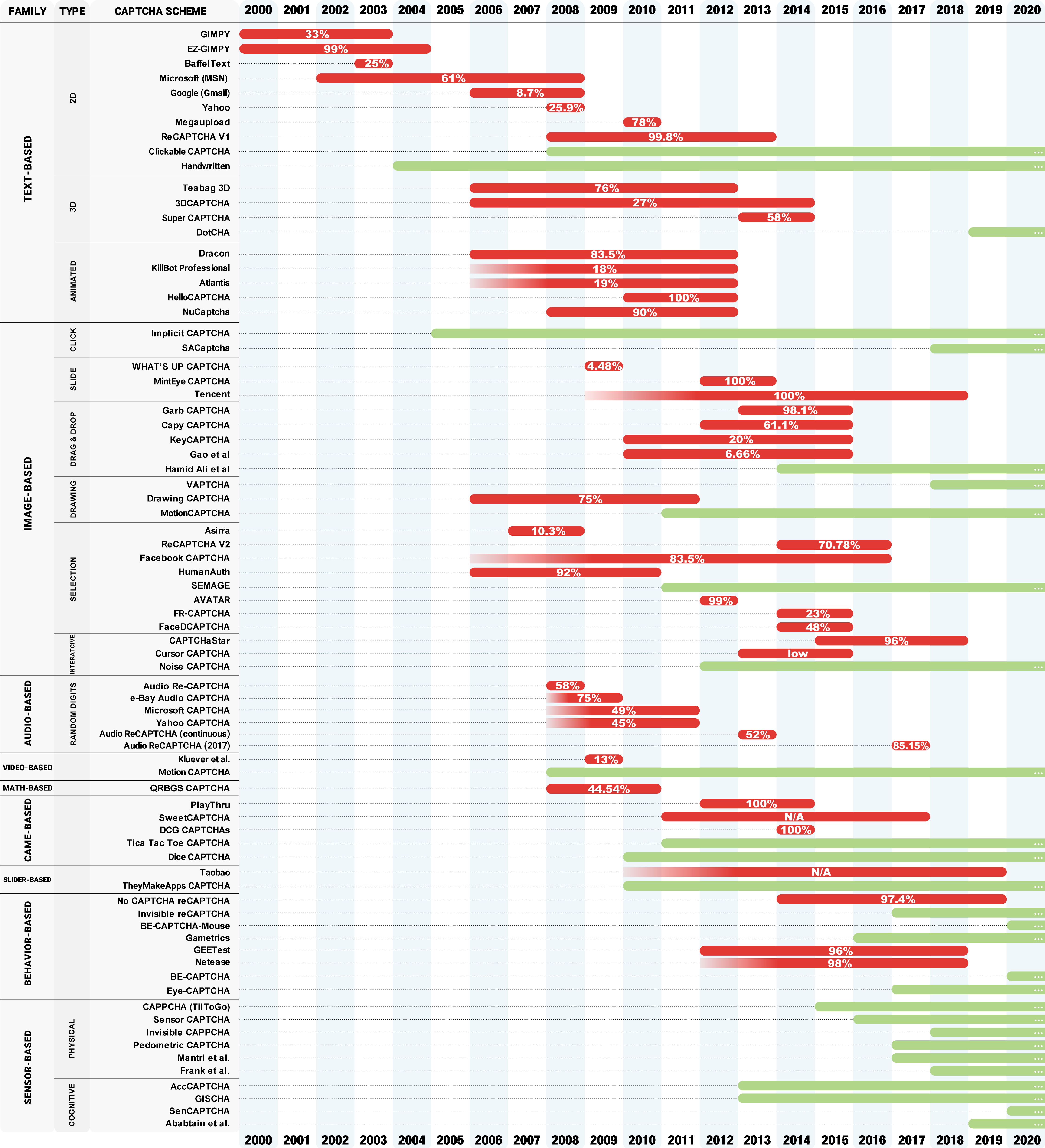}
  \caption{Timeline of the security breaking of the CAPTCHA schemes}
  \label{fig:timeline}
\end{figure*}

In the following, we outline the existing techniques for attacking the different CAPTCHA schemes presented in section 2. Furthermore, we plot them in a timeline graph (Figure \ref{fig:timeline}) to report if the scheme has been broken (represented in the graph with a red bar), the number of years that occurred to find a successful attack, and the best breaking percentage achieved. As shown in Figure \ref{fig:timeline}, most CAPTCHA schemes have been successfully broken with a high success rate in few years.

\subsection{Attacks against text-based CAPTCHA}

A lot of works suggested methods to break the different type of text-based CAPTCHAs. 
In 2003, Mori and Malik \cite{ref10} proposed a method based on shape context matching to break both Gimpy and EZ-Gimpy CAPTCHAs with 33\% and 92\% accuracy, respectively. 
In \cite{ref11}, EZ-Gimpy was also broken with a success rate of 99\% using a correlation algorithm and a direct distortion estimation algorithm. 
In 2005, Chellapilla et al. \cite{ref14}, \cite{ref15} were able to break various text-based CAPTCHAs by using machine learning, and suggested a secure CAPTCHA scheme based on hard-segmentation problems. 
In 2008, Yan and El Ahmad showed that some segmentation-resistant CAPTCHAs could be broken, including the ones used by Microsoft, Google, and Yahoo \cite{ref16}, \cite{ref17}. %However, the success rate of Google CAPTCHA was only 8.7\%. 
%\textcolor{red}{
Later, other researchers attempt also to break these CAPTCHA schemes and they achieved higher success rates \cite{ref152},\cite{ref145}. El Ahmad and Yan \cite{ref18} were able to break Megaupload CAPTCHA with a success rate of 78\%. In 2014, researchers from Google \cite{ref20} broke the hardest category of ReCAPTCHA using neural networks with an accuracy of 99.8\%.

In \cite{ref26}, the authors discovered a set of attacks against 3D CAPTCHAs, even without the usage of OCR programs. In details, they were able to successfully extract a set of pixels from the characters of several 3D CAPTCHA schemes (i.e., Teabag 3D, 3dcaptcha, and Super CAPTCHA) that can be used for automated recognition of the challenge. Thanks to such a technique, the authors were able to achieve success rates of 31\%, 58\%, and 27\% in breaking Teabag 3D, 3dcaptcha, and Super CAPTCHA, respectively.
%In \cite{ref26}, the authors stated that the fact that OCR programs cannot be used directly to solve the 3D CAPTCHAs do not make them more secure than their 2D counterparts. They demonstrated this by showing how certain patterns in 3D CAPTCHAs ( i.e., Teabag 3D, 3dcaptcha, and Super CAPTCHA) can be exploited to extract pixels that belong to the characters. Once this information is extracted, automated attacks were used to solve these 3D CAPTCHAs with a high success rate. They achieved success rates of 31\%, 58\%, and 27\% to break Teabag 3D, 3dcaptcha, and Super CAPTCHA, respectively. 
Furthermore, the same authors in \cite{ref25} were able to break Teabag 3D with a higher success rate (i.e., 76\% ) by exploiting the side surface information contained in the 3D text objects. 
%They used this information to separate the text from the background as well as to segment the characters. 

%% I don't think is relevant
%In \cite{ref27}, chow et al. stated that the proposed 3D CAPTCHA by Imsamai and Phimoltares \cite{ref24} is not secure against automated attacks, and this is mainly due to the fact that the authors didn't take into consideration the advances in computer vision on 3D object recognition.

%Designers of animated CAPTCHAs assumed that this type of CAPTCHA provides better security over traditional single image text-based CAPTCHAs because important information can be spread over multiple animation frames, rather than being contained within a single image. 
%However, 
Nguyen et al. \cite{ref33} showed that the information across multiple animation frames in animated CAPTCHA schemes could be easily extracted using simple techniques such as the PDM (Pixel Delay Map) or CL (Catching Line) methods. 
They used these methods to defeat several animated CAPTCHAs with a high success rate, including iCAPTCHA, Atlantis, KillBot Professional, and Dracon CAPTCHA. 
%Therefore, adding the time dimension alone is not enough to make animated CAPTCHA more secure. 
In \cite{ref34}, the same methods have been used to defeat different types of HelloCAPTCHA schemes with a success rate between 16\% -100\%, due to their weakness against segmentation attacks.  
%This is mainly due to the fact that HelloCAPTCHA was not designed with the segmentation resistant principle in mind. 
Unlike HelloCAPTCHA, NuCaptcha is an animated CAPTCHA designed to be segmentation resistant. Since the characters are overlapped and crowded together, the PDM or CL methods used to defeat HelloCAPTCHA are not effective to separate the characters. 
However, NuCaptcha has been broken using more sophisticated attacks \cite{ref35}, \cite{ref36}. Elie Bursztein \cite{ref36} achieved a success rate of 90\% by using bounding box shape analysis and an interest points (SIFT algorithm) density evaluation to isolate objects in each frame. Then he tracked these objects across multiple frames and kept only the 50 frames that contain the CAPTCHA answer.

\subsection{Attacks against image-based CAPTCHA}

Many attacks have been suggested in the literature to bypass the different type of image-based CAPTCHAs. Golle \cite{ref40} was able to break the Asirra scheme with a success rate of 10.3\%. To do so, he used different features to train an SVM (Support Vector Machine) classifier to identify cats and dogs with 82.7\% accuracy (i.e., accuracy for a single image). 
Hernandez-Castro et al. in \cite{ref48} proposed a side-channel attack that bypassed the HumanAuth challenge with a 92\% success rate. 
Sivakorn et al. \cite{ref41} have successfully attacked both Google and Facebook image-based CAPTCHA with success rates of 70.78\% and 83.5\%, respectively. 
In \cite{ref42}, the authors broke the new and the old variation of reCAPTCHA V2 with 79\% and 88\% success rates, respectively. 
Furthermore, they broke the Facebook image CAPTCHA and the China Railway CAPTCHA with success rates of 86\% and 90\%, respectively. 
Cheung \cite{ref44} successfully broke Avatar CAPTCHA using Convolutional Neural Networks (CNN), with a high success rate of 99\%. Gao et al. \cite{ref47} broke both FR-CAPTCHA and FaceDCAPTCHA with success rates of 23\% and 48\%, respectively.

%Minteye  CAPTCHA  was  broken  by  a  very  simple  attack  based  on  Sobel  operators  that only  required  23  lines  of  Python. 

The Minteye CAPTCHA scheme was broken in \cite{ref51}, by exploiting the concept of Sobel operators and the length of the edges of the image.
The idea behind this attack is based on the observation that the more an image is swirled, the longer the edges in the image become. So, the breaking methods consists in summing the length of the edges in the image and then select the image with the lowest sum of edges as the correct answer. 

In \cite{ref42}, the authors broke different schemes of image-based CAPTCHAs, including the Tencent CAPTCHA. In detail, their proposal achieved 100\% success rate even during the motion of the sliding puzzle to the target region.
%, which leads them to deduce that this CAPTCHA does not use any mechanism for detecting malicious behaviors.
Hernandez-Castro et al. \cite{ref58} proposed a very low-cost attack that does not attempt to solve image recognition or shape recognition problems but instead uses JPEG to measure the continuity of the image. Through this side-channel attack, they were able to bypass the most popular sliding-based CAPTCHAs. In detail, they break Capy CAPTCHA with a 65.1\% success rate, and by applying minor modifications, they were able to break KeyCAPTCHA and Garb CAPTCHA as well with success rates of 20\% and 98.1\%, respectively. 
Conti el al \cite{ref59} pointed out that Jigsaw CAPTCHA proposed by Gao et al. \cite{ref56} is vulnerable to relay attack and random guess attack with a success rate of 6.66\%.
Lin et al. \cite{ref63} broke Drawing CAPTCHA with an accuracy of 75\%. They proposed an effective erosion-based breaking algorithm based on their observation of the difference between the size of the diamond-shaped dots and the dots used in the background as noise. 
%The other drawing-based CAPTCHAs did not get much attention in the literature.

Although CAPTCHaStar authors tested its resiliency against several types of automated attacks such as traditional attacks, automated attacks using ad-hoc heuristics, and attacks based on machine learning, recently, Gougeon and Lacharme \cite{ref67} were able to break this CAPTCHA with a 96\% success rate. In addition, they pointed out that the modiﬁcation of the parameters does not prevent CAPTCHaStar against their proposed attack, which is based on the concentration of pixels (i.e., stars) during the formation of the image. In \cite{ref59} the authors pointed out that the resiliency of Cursor CAPTCHA to machine learning-based attacks and stream relay attack is low.

\subsection{Attacks against audio-based CAPTCHA}

Tam et al. \cite{ref73} were the first to evaluate the robustness of audio CAPTCHAs against automated attacks. They were able to break audio reCAPTCHA using an SVM-based approach. They achieved a success rate of 45\% when they matched the solution exactly and 58\% when they leveraged a ``one mistake'' passing condition. 
Burzstein and Bethard \cite{ref74} introduced Decaptcha, a system that was able to bypass the eBay's audio CAPTCHAs with a 75\% success rate. Their system applies a Discrete Fourier Transform (DFT) to the wave file and then isolates the energy spikes. Afterward, it uses a supervised learning algorithm to recognize speech patterns. 
In \cite{ref76} the authors proposed a CAPTCHA solver based on the non-continuous speech, which defeated the Microsoft and the Yahoo audio CAPTCHAs with a success rate of 49\% and 45\%, respectively. The segmentation phase was unsupervised, while the classification phase was supervised. They used the Regularized Least-Squares Classification (RLSC) algorithm for classification, and Amazon Mechanical Turk to label scraped CAPTCHAs. 
However, their system was able to solve reCAPTCHA with only 1.5\% success rate, due to the presence of semantic vocal noise. 
Sano et al. \cite{ref75} developed a CAPTCHA solver for continuous CAPTCHAs that use overlapping target voices as defensive techniques to make automated segmentation difficult. Their system applied Hidden Markov Models (HMMs) for speech recognition. It was tested on the version of audio reCAPTCHA used in 2013, and the results show that it was able to break this version of continuous reCAPTCHA with a success rate of 52\%. 
Bock et al. \cite{ref77} introduced unCaptcha, an automated system that can bypass audio reCAPTCHA released in 2017 with an 85.15\% success rate. They attained these results by leveraging free online speech-to-text services and performing a minimal phonetic mapping to enhance accuracy.

\subsection{Attacks against Behavior-based CAPTCHA}

Although Sliding-based behavioral CAPTCHA schemes attempted to increase the security of sliding CAPTCHAs by detecting malicious behaviors, recently, Zhao et al. \cite{ref42} were able to bypass such a detection by leveraging four simulation functions (i.e., Sigmoid, Softmax, ReLu, and Tanh) to mimic human behaviors. 
Their proposed attack against the GeeTest and Netease CAPTCHA schemes achieves the best success rate of 96\% and 98\% respectively, by using the Sigmoid function. 
Furthermore, Sivakorn et al. \cite{ref41} found that Google's tracking cookies can be used to influence the risk analysis and, thus, bypass the reCAPTCHA V2 restrictions. In detail, the authors designed a tracking cookie for bots that was able, after nine days of automated browsing across different Google services, to fool the Google risk analysis system into thinking that the traffic is made by human beings and, consequently, to check the ``I'm not a robot'' box. Furthermore, the authors proposed a low-cost attack that breaks the second layer of reCAPTCHA V2 with a success rate of 70.78\%. 
In \cite{ref110}, the authors used a ``divide and conquer'' strategy to defeat the No CAPTCHA reCAPTCHA scheme for any grid resolution. They achieved a success rate of 97.4\% on a {100 x 100} grid and 96.7\% on a {1000 x 1000} screen resolution.

\subsection{Attacks against the other type of CAPTCHA}

%Besides video-based CAPTCHAs which did not receive much attention, in the following 
%In this section, we present some attacks that have been suggested to compromise the other type of CAPTCHAs. 

Kluever el al. \cite{ref85} performed a tag frequency-based attack to evaluate the security of their proposed video-based CAPTCHA and achieved a success rate of 13\%. Hernandez-Castro el al. \cite{ref109} were able to break QRBGS CAPTCHA using a side-channel attack with a success rate of 44.54\%. In \cite{ref80}, Mohamed et al. reported that DCG CAPTCHAs, including PlayThru, are vulnerable to dictionary-based automated attacks. 
In \cite{ref83}, a developer proposed a solver that automatically bypasses SweetCAPTCHA. 
In \cite{ref52},  different variations of slider CAPTCHAs, including the Taobao scheme, have been bypassed by using a simple JavaScript code and puppeteer.

\begin{comment}
\textcolor{red}{
\subsection{Overall Discussion}
As shown in Figure \ref{fig:timeline}, most CAPTCHA schemes have been successfully broken with a high success rate in few years.
Indeed, the recent advancements of AI technology drove the exploitation of multiple attacks to both traditional and behavioral-based CAPTCHAs, including the No CAPTCHA reCAPTCHA scheme (with a 97.4\% of success rate). 
On the other hand, Sensor-based CAPTCHAs, especially the schemes based on the physical nature of humans, seem promising and more resilient to attacks as there are no publicly available attacks to those schemes.
Still, further research is needed to investigate the security of those type of CAPTCHAs, as described in Section \ref{sec:challenge_security}}.
\end{comment}

\section{Evolution of CAPTCHA Schemes}
%\section{Discussion on the current state-of-the-art of CAPTCHA}
\label{sec:discussion}

%%% text-based captcha

The evolution of CAPTCHA schemes follows the advancements of technology to break them.
In the early 2000s, text-based CAPTCHAs were the dominant solutions to discern between human and automated users. To this aim, security experts developed a set of attacks to break the most popular text-based schemes by leveraging image processing, pattern recognition, and machine learning algorithms \cite{ref111}.
Furthermore, the scientific community attempted to enhance the security of existing text-based CAPTCHAs by applying anti-segmentation and anti-recognition techniques. However, these countermeasures made text-based CAPTCHAs challenging even for human users, resulting in a higher error rate and limited usability that reduces text-based schemes' popularity.
Finally, in 2014 a research conducted by Google demonstrated that the advancements in the AI technology could solve the most complicated variants of distorted text at 99.8\% accuracy \cite{ref20}, leading to the decline of the text-based CAPTCHA schemes.

%%% image-based captcha

The security weaknesses of text-based CAPTCHAs and its usability issues, especially with the advent of mobile devices, led many researchers to look for alternatives. Since 2004, many of them have focused on exploiting Computer Vision (CV) problems such as image classification and object recognition that were considered harder AI problems than character recognition at that time. 
Chew and Tygar \cite{ref112} were among the first researchers using labeled images to design image-based CAPTCHAs. After that, many images-based CAPTCHAs schemes have been proposed to create challenges that require selection, drag and drop or sliding of images to discern between human and automated usages. However, the advancement in CV and machine learning and the advent of Machine Learning as a service (MLaaS) solutions boosted the breaking of the major image-based CAPTCHA schemes between 2013 and 2018.
%In detail, the study conducted by Zarras et al. \cite{ref117} reveals that MLaaS services do not require any machine learning expertise, the availability of large datasets for training, or the operation of the necessary ML infrastructures.
For instance, the authors of \cite{ref42} exploited ML to perform attacks against several image-based CAPTCHAs, including the image-based reCAPTCHA V2 scheme. 
%\textcolor{red}{
Furthermore, the authors proposed several countermeasures, including the use of distortion techniques on characters on the background image or in the hint, the addition of noise on background images, and the use of adversarial examples to hinder deep learning models.
In this regards, the concept of adversarial examples was first introduced by Szegedy et al.\cite{ref154}, and since then, many researchers proposed CAPTCHA schemes based on adversarial examples to improve its security against ML-based bot attacks \cite{ref155},\cite{ref156},\cite{ref157}. 
However, Na et al.\cite{ref158} recently proposed an efficient CAPTCHA solver that breaks adversarial CAPTCHAs using incremental learning with only a small dataset. The authors demonstrated that existing defense methods (e.g., \cite{ref156}, and \cite{ref157}) that use adversarial examples in CAPTCHA schemes are not effective against their proposed adaptive CAPTCHA solver. %}
%\textcolor{red}{The authors proposed as well some countermeasures which include applying distortion techniques on characters on the background image or in the hint, adding noises on background images or using adversarial examples which are imperceptible to humans, yet can fool deep learning models.
%In fact, the concept of adversarial examples was first introduced by Szegedy et al.\cite{ref154} and since then many researchers proposed CAPTCHA schemes based on adversarial examples to improve its security against ML-based bot attacks \cite{ref155},\cite{ref156},\cite{ref157}. However, Na et al.\cite{ref158} recently proposed an efficient CAPTCHA solver that breaks adversarial CAPTCHAs using incremental learning with only small dataset. The authors demonstrated that existing defense methods \cite{ref156},\cite{ref157}, that use adversarial examples in CAPTCHA schemes, are not effective to defend against their proposed adaptive CAPTCHA solver.}

%%%%% audo-based captcha

In conjunction with the advent of text-based and image-based CAPTCHAS, the security experts proposed Audio-based CAPTCHAs to cope with visually impaired users. However, those schemes are limited by language barriers and low usability, as discussed in \cite{ref113}. Furthermore, they are also weak against supervised learning, and automatic Speech Recognition (ASR) attacks \cite{ref114}.

Starting from the 2010s, the research community introduced behavioral-based CAPTCHA schemes to build challenges based on behavioral biometrics measurements. The first deployed behavioral-based CAPTCHA was introduced in 2012 by the Geetest company, while in 2014, Google released No CAPTCHA reCAPTCHA and later on Invisible CAPTCHA (2017).
%Recently many researchers also focused on behavioral CAPTCHA to enhance the security of conventional CAPTCHAs by adding a transparent security layer to detect malicious behaviors rather than increasing the complexity of the challenge. 
Still, most of the commercial and academic proposals are based on mouse dynamics, which have been shown to be vulnerable to bots attacks that attempt to mimic the user's behavioral pattern \cite{ref42}, \cite{ref137}. 
As shown in the timeline of Figure. \ref{fig:timeline}, the most widespread behavioral CAPTCHAs (i.e., No CAPTCHA reCAPTCHA, GEETest, and Netease) have been broken with a high success rate \cite{ref110}, \cite{ref42} in the last years. 

%More research work is needed to investigate the security of Invisible reCAPTCHA and the other academic proposals which have not been broken yet. 
In addition, behavioral-based CAPTCHA schemes raise serious privacy concerns as described in \cite{ref120}, \cite{ref116}, and \cite{ref92}. 
For instance, \cite{ref116} demonstrated how demographic attributes such as gender, age group, and education level could be extracted while solving a simple game CAPTCHA (e.g., Gametrics) by capturing user's innate cognitive abilities and behavioral patterns. 
Due to such concerns, Cloudflare recently decided to move away from reCAPTCHA \cite{ref119}. 

Finally, the latest research directions exploit the data gathered from sensors to build challenges that are difficult to be emulated by automated bots.
At the time of writing, no study has been done to review or analyze the security strength of sensor-based CAPTCHAs, and none of the proposed solutions has been successfully bypassed.
%The main limitation of CAPTCHAs in this category is the fact that it is not functional in devices not enabled with the required sensors.

%Until the time of writing this paper, no study has been done to review or analyze the security strength of CAPTCHAs based on sensor readings and none of the proposed solution has been cracked yet.
%To the best of our knowledge, the first company that implemented a proof of concept of sensor-based CAPTCHA and attempted to deploy it is Brave \cite{ref97}. 
%The main limitation of CAPTCHAs in this category is the fact that it is not functional in devices not enabled with the required sensors.
%s\footnote{a full rewrite is needed, but the concept is worth expressing} 

\section{Open Issues, Challenges and opportunities}\label{sec:challenge}

%Although the comparison between the different CAPTCHA categories is not straightforward because not all CAPTCHA schemes from the same category necessarily share the same advantages and limitations. Nevertheless, in this section we provide a comparison between the common CAPTCHA groups from security, usability, privacy and compatibility standpoint, taking in consideration the majority. We believe it would be beneficial for researchers to have an overview on the advantages and the limitations of each group from different aspects. 

%In this section we compare between the different CAPTCHA mechanisms from security, usability, privacy and compatibility standpoint. 

In this section, we identify the open issues in designing robust and usable CAPTCHA schemes, as well as the main challenges that a CAPTCHA designer might have to deal with, and opportunities for further study.% In addition, we identify some potential aspects that are worth further study.%\footnote{you are worth something you don't worth something}.

\subsection{Resilience to both automated and human solver relay attacks}
\label{sec:challenge_security}
A CAPTCHA scheme can be considered highly secure when both the automated attack success rate is less than 0.01\% \cite{ref16},\cite{ref142} and it is resilient to human solver relay attacks. Unfortunately, in the literature most studies dedicated to the design of CAPTCHA schemes focus only on automated attacks while only few of them take into account the resilience to human solver relay attacks. 

The security level of traditional CAPTCHA schemes depends on the hardness of some AI problem. However, the progress of AI techniques and computing power has led to the breaking of these CAPTCHA schemes with high success rates \cite{ref20},\cite{ref42}, \cite{ref41},\cite{ref77}. Therefore, in order to design the next generation CAPTCHA schemes, it is important to move away from schemes based on hard AI problems toward other approaches less vulnerable to learning-based attacks \cite{ref115}. Recently, big companies like Google, Alibaba and Tencent have migrated towards behavior-based CAPTCHA schemes, while there is an initiative aiming at deploying a sensor-based CAPTCHA scheme that uses the same key concept of Invisible CAPPCHA \cite{ref97} by a company called Brave \cite{ref116}.

As presented in detail in Section 3, all the popular conventional CAPTCHA schemes have been broken with high success rate by automated attacks and most of them are also vulnerable to human solver relay attacks (the most notable exceptions being CAPTCHaStar, PlayThru and Dynamic Cognitive Game CAPTCHA). Similarly, popular behavior-based CAPTCHA schemes have also been broken with high success rate by automated attacks, and all of them are vulnerable to human solver attacks. Invisible reCAPTCHA and other academic proposals have not been broken yet, however with the advent of the fourth generation bots which rotate through thousands of different IP addresses and mimic accurately the human behavior, it would be difficult to design a secure CAPTCHA based solely on the user behavior data that can be gathered in a normal (i.e., with no additional sensors or special hardware) environment. None of the sensor-based CAPTCHA has been broken yet by automated attacks, however similar to the other types of CAPTCHA schemes, most of them are vulnerable to human solver relay attacks. The exception to this vulnerability is represented by the ones that have been specifically designed to resist this kind of attack (e.g., Invisible CAPPCHA). Another weakness of sensor-based CAPTCHA schemes is the limited number of challenges. This is due to the fact that designing a large number of usable gestures for instance, to ensure high security against automated attacks, is very challenging. However, this weakness may be solved relying on trusted hardware.

%\footnote{this is a DOH! point: "design something that works"}.

On the basis of the above observations, we identified the following open problems %potential aspects\footnote{let's say "open problems" instead} 
that require further study in order to design robust and usable CAPTCHA schemes:  it is necessary to investigate 1) the resilience of currently unbroken behavior-based CAPTCHAs against fourth generation bots; 2) the security strength of sensor-based CAPTCHA schemes against replay attacks, sensor manipulation \cite{ref143} and human solver relay attacks 3) the security of CAPTCHA schemes that make validation process at the client-side either with or without secure hardware as they may be hacked.

\subsection{Friction-heavy vs. Frictionless Challenges}

CAPTCHA schemes are well known as a source of annoyance to users. This is due to the fact that most of the time designers trying to make the scheme more secure also make the challenge harder for humans. It is important to reduce the friction in general and the cognitive overload associated to the challenges. Creating user-friendly CAPTCHAs, yet, it is not always an easy task and in many cases there is a trade-off between security and usability. Some CAPTCHA schemes achieve complete transparency to users (i.e., invisible reCAPTCHA, invisible CAPPCHA) removing all cognitive challenges. However, it is worth noting that not all the CAPTCHA schemes in the same category (i.e., behavioral-based and sensor-based) are automatically endowed with the same level of usability. In fact, while some of them require a simple task such as clicking on a check box or tilting the device, others requires less user-friendly tasks such as solving a complex cognitive task, performing a physical task such as walking few steps or performing complex gestures.

%Another category of CAPTCHA schemes that is generally user-friendly is sliding-based CAPTCHAs, especially the ones that do not require image recognition. followed by game-based CAPTCHA and image-based CAPTCHAs. Text-based, audio-based, video-based, math-based CAPTCHAs are considered the less usable CAPTCHAs due to the high error rate, in addition that they are not pleasant. 

To the best of our knowledge, there is no study fully dedicated to the analysis of the usability of behavior-based and sensor-based CAPTCHA schemes. Therefore, we argue that such a study would allow assessing the level of usability of all the CAPTCHA schemes proposed in the behavioral-based and sensor-based categories.

\subsection{Preserving the user's privacy}
 
Unlike traditional CAPTCHA schemes, it has been shown that the new behavior-based and sensor-based CAPTCHA schemes may raise a privacy issue when information such as user's behavioral data, sensor data and cookies that can be used for tracking are sent to a remote server. As a solution, some researchers suggested to send solely the results of the test to the server, instead of the sensor data. However, trusted hardware is then required to prevent hacking at the client side. Further study is needed to identify methodologies capable of preventing client-side hacking without requiring trusted hardware. Besides, the user's privacy should be taken into strong consideration in general from the very start of the design phase of new CAPTCHA schemes. 

\subsection{Compatibility with all devices}

A robust and usable CAPTCHA scheme that is compatible with different form factors is obviously highly desirable, however, the most promising CAPTCHA schemes category in term of security and usability present a significant dependency on a specific form factor. For instance, behavioral-based CAPTCHA schemes strongly rely on mouse dynamics or on touch-and-tap dynamics, hence they require form-factor specialization. Sensor-based CAPTCHA schemes require sensors that are available only in tablets, smartphones and smartwatches (e.g., \cite{ref138}, \cite{ref141}), hence they are currently unavailable on a large portion of users' devices and further study to find potential surrogates of sensors data, possibly relying on trusted hardware, on desktops and laptops are needed.

\section{Conclusion
\label{sec:conclusion}}%\footnote{redo: 1) it's conclusions, so we have done, not we aim at doing 2) a bit of what we wrote in security analysis and open issues will be nice here}

CAPTCHA has been widely used as a security mechanism to prevent bots from abusing online services. Over the years, different types of CAPTCHA schemes have been proposed, mainly to improve the usability and the security against new threats presented by evolving bots. The studies in the literature usually focus on the conventional CAPTCHA schemes, i.e., text, image and audio based schemes, and do not take into account either new types of schemes or novel threats such as human solver relay attacks, sensor manipulation \cite{ref143} and the risk of privacy breaches. In this paper, we have provided a comprehensive review of the related research involving two decades, by also highlighting the new trends and open issues. We have first presented a comprehensive classification of the current CAPTCHA schemes that includes both traditional and new ones. Then, to evaluate their drawbacks from the security point of view, we have provided a detailed summary on the attack methods that have been used to break CAPTCHA schemes in each category. Finally, we have discussed the current state-of-the-art in the field of CAPTCHA schemes design, highlighting the open issues, the challenges, and the opportunities for further research that constitute the road toward the design of the next generation of secure and user-friendly CAPTCHA schemes.
%\begin{acks}
%To Robert, for the bagels and explaining CMYK and color spaces.
%\end{acks}

\bibliographystyle{ACM-Reference-Format}
\bibliography{biblio}

%\appendix

\end{document}

%% file: tables/table1.tex
%%%%%%%%%%%%%%%%%%%%%%% TABLE 1 GOES HERE %%%%%%%%%%%%%%%%%%%%%%%%

%\onecolumn
%\afterpage{
%\FloatBarrier
%\begin{table*}[h!]
%\begin{table}[H]
%\scriptsize
%\centering

% with vertical tables
%
{
\scriptsize
\centering
\begin{longtable}[]{m{0.08\textwidth} m{0.18\textwidth} m{0.26\textwidth} m{0.03\textwidth} m{0.30\textwidth}}

%\twocolumn
%without vertical lines
%\begin{tabular}[]{m{0.08\textwidth}m{0.16\textwidth}m{0.20\textwidth}m{0.03\textwidth}m{0.4\textwidth}}

\hline  \textbf{Type} & \textbf{Scheme} & \textbf{Sample} & \textbf{Year} & \textbf{Challenge Description} \\ \toprule

%\endhead

 \multirow{10}{*}[-120pt]{2D} & GIMPY \cite{ref9} & \includegraphics[width=\imgW, height=1.5cm]{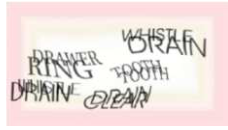} & 2000 & Recognize three words out of of seven selected randomly from a dictionary \\ 
\cmidrule{2-5} 

  & EZ-GIMPY  \cite{ref9} &  \includegraphics[width=\imgW, height=\imgT]{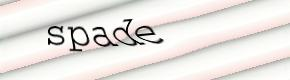} & 2000 & Recognize one English word in a distorted image \\ 
 \cmidrule{2-5} 

  & BaffelText \cite{ref12}  & \includegraphics[width=\imgW, height=\imgT]{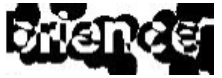} & 2003 & Recognize a pronounceable string of characters with difference masking applied \\ 
 \cmidrule{2-5} 

  & Microsoft (MSN) \cite{ref16} &  \includegraphics[width=\imgW, height=\imgT]{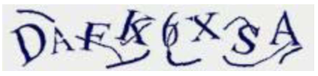} & 2002 & Recognize eight distorted characters presented with random arcs as clutters \\ \cmidrule{2-5} 
 
  & Google (Gmail) \cite{ref16}     &  \includegraphics[width=\imgW, height=\imgT]{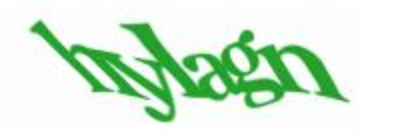}      & 2006 & Recognize characters which are  crowded together \\ 
 \cmidrule{2-5} 

  & Yahoo  \cite{ref16} & \includegraphics[width=\imgW, height=\imgT]{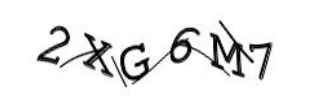} & 2008 & Recognize a string of characters connected by intersecting random lines \\ \cmidrule{2-5} 
 
  & Megaupload &  \includegraphics[width=\imgW, height=\imgT]{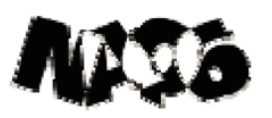} & 2010 & Recognize four overlapped characters with negative intersection areas \\ \cmidrule{2-5}
 
  & ReCAPTCHA V1  \cite{ref19} & \includegraphics[width=\imgW, height=\imgT]{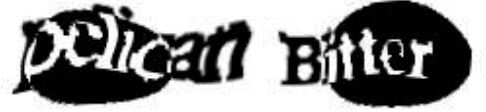} & 2008 & Recognize distorted text scanned from old books \\ 
 \cmidrule{2-5} 

  & Clickable CAPTCHA \cite{ref21} &  \includegraphics[width=\imgW, height=\imgT]{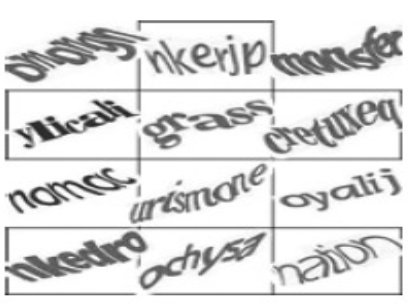} & 2008 & Identify English words among non-English words \\ 
 \cmidrule{2-5}
 
  & Handwritten \cite{ref22} & \includegraphics[width=\imgW, height=\imgT]{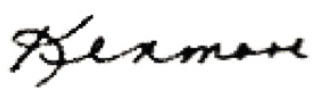} & 2004 & Recognize a distorted handwritten text (e.g., city name) \\ 
 \cmidrule{1-5}
 
  \multirow{4}{*}[-40pt]{3D} & Teabag  3D \cite{ref7} & \includegraphics[width=\imgW, height=\imgT]{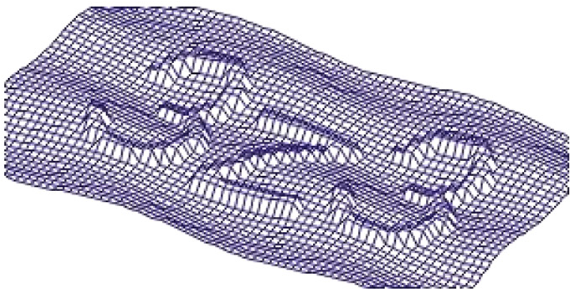}   & 2006 & Recognize a sequence of characters that appears on a grid in 3D space \\ 
 \cmidrule{2-5} 
 
  & 3DCAPTCHA \cite{ref26} & \includegraphics[width=\imgW, height=\imgT]{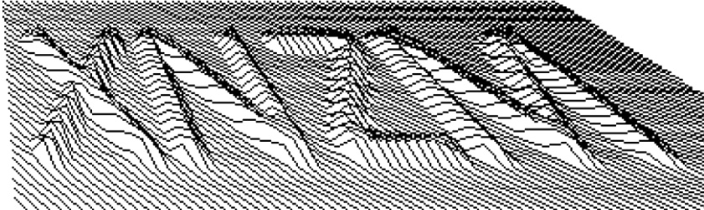} & 2006 & Recognize a sequence of 3D characters\\ 
 \cmidrule{2-5}
 
  & Super CAPTCHA \cite{ref8} & \includegraphics[width=\imgW, height=\imgT]{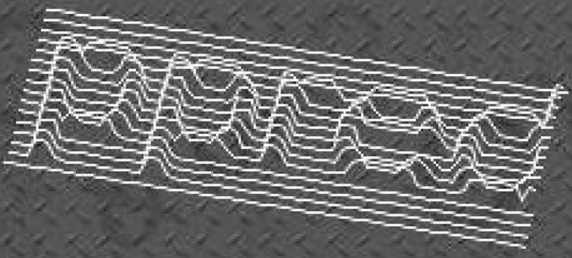} & 2013 & Recognize a sequence of 3D characters \\ 
 \cmidrule{2-5} 

 & DotCHA \cite{ref28} & \includegraphics[width=\imgW, height=\imgT]{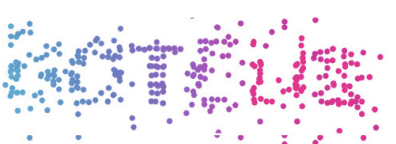} & 2019 & Drag and rotate the model to identify each letter, then type the answer \\ 
\cmidrule{1-5}

 \multirow{5}{*}[-40pt]{Animated} & Dracon CAPTCHA \cite{ref6} & \includegraphics[width=\imgW, height=\imgT]{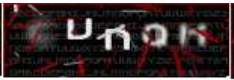} & 2006 & Recognize five characters which fade and blur at various times over the animation frames  \\ \cmidrule{2-5} 

  & KillBot Professional \cite{ref33} &  \includegraphics[width=\imgW, height=\imgT]{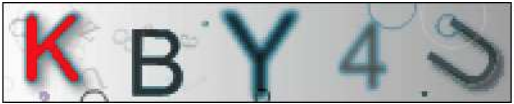} &     & Recognize five moving characters among a noisy foreground and/or background \\ 
 \cmidrule{2-5}
 
 & Atlantis CAPTCHA \cite{ref33} & \includegraphics[width=\imgW, height=\imgT]{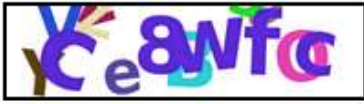} &  & Recognize six moving characters among other continuously changing their color \\ 
 \cmidrule{2-5} 

  & HelloCAPTCHA \cite{ref32} &  \includegraphics[width=\imgW, height=\imgT]{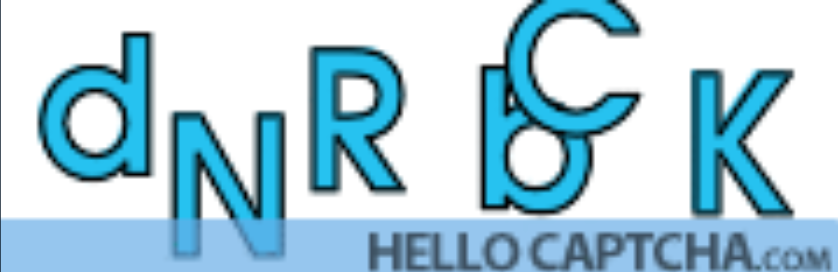} & 2010 & Recognize a sequence of six characters displayed in an animated GIF image \\ 
 \cmidrule{2-5}
 
  & NuCaptcha \cite{ref37} & \includegraphics[width=\imgW, height=\imgT]{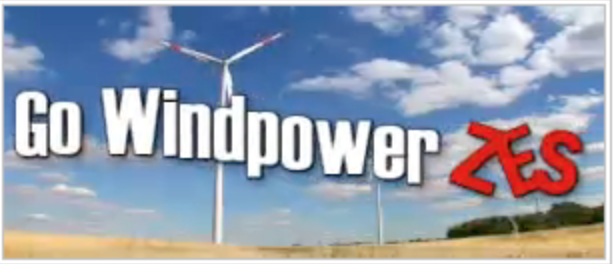} & 2008 & Type the last three red moving characters \\ 
 \midrule

%\end{tabular}
\caption{A taxonomy of text-based CAPTCHAs }
\label{table:textCAPTCHA}

%\end{table*}
\end{longtable}
%\FloatBarrier
%\end{table}
}

%% file: tables/table2.tex
%\onecolumn
\afterpage{
\scriptsize
\centering

% with vertical tables
%\begin{longtable}[]{|m{0.13\textwidth}|m{0.11\textwidth}|m{0.16\textwidth}|m{0.18\textwidth}|m{0.03\textwidth}|m{0.22\textwidth}|}
%\twocolumn
%without vertical lines
\begin{longtable}[]{m{0.10\textwidth}m{0.24\textwidth}m{0.20\textwidth}m{0.03\textwidth}m{0.28\textwidth}}

\hline \textbf{Type} & \textbf{Scheme} & \textbf{Sample} & \textbf{Year} & \textbf{Challenge Description} \\ \toprule

\endhead

%%%%%%%%%%%%%%%%%%%%%%%%%%%%%%%%%

% Image-based Click

  \multirow{2}{*}[-28pt]{Click} & Implicit CAPTCHA \cite{ref60} & \includegraphics[width=\imgW, height=\imgL]{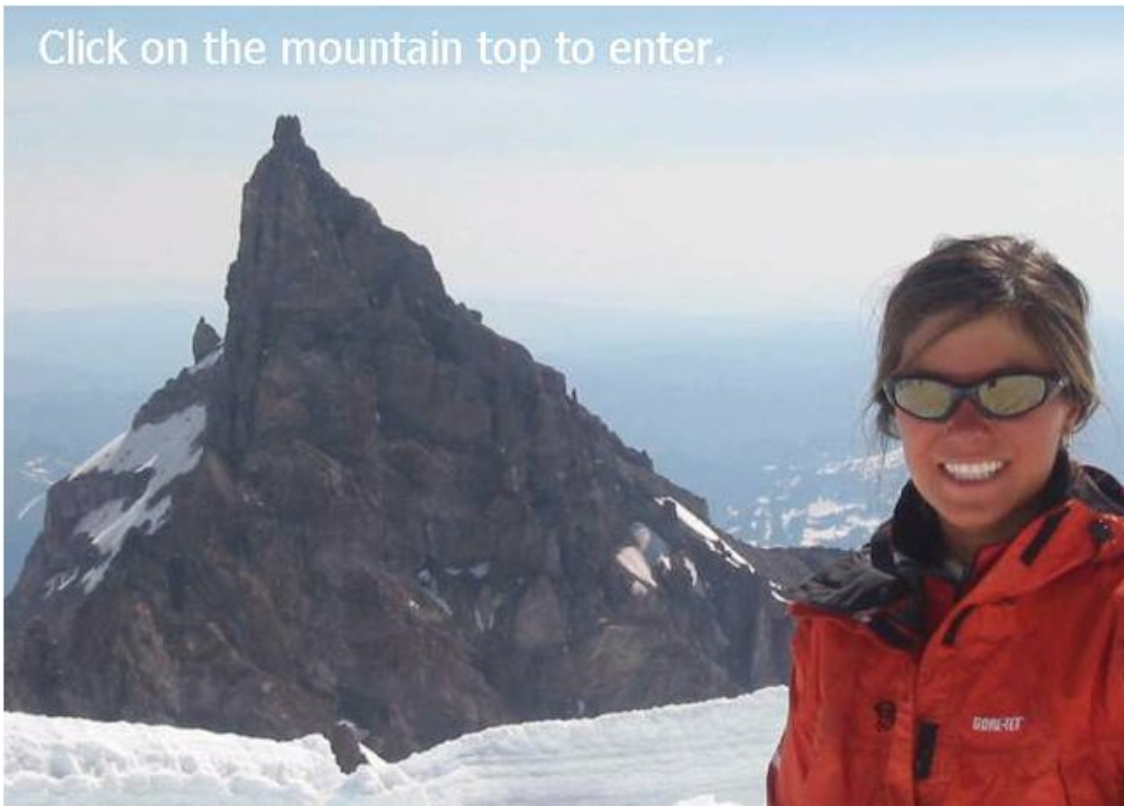} & 2005 & Click on a specific area of an image (e.g., mountain top) \\ 
\cmidrule{2-5}

 & SACaptcha \cite{ref61} & \includegraphics[width=\imgW, height=\imgL]{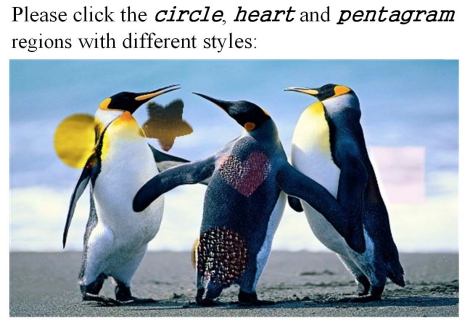} & 2018 & Click on some regions in the image that have a specific shape mentioned in the challenge description \\ 
\cmidrule{1-5} 

% Image-based Slide

 \multirow{3}{*}[-56pt]{Sliding} &  WHAT'S UP CAPTCHA \cite{ref50} & \includegraphics[width=\imgW, height=\imgL]{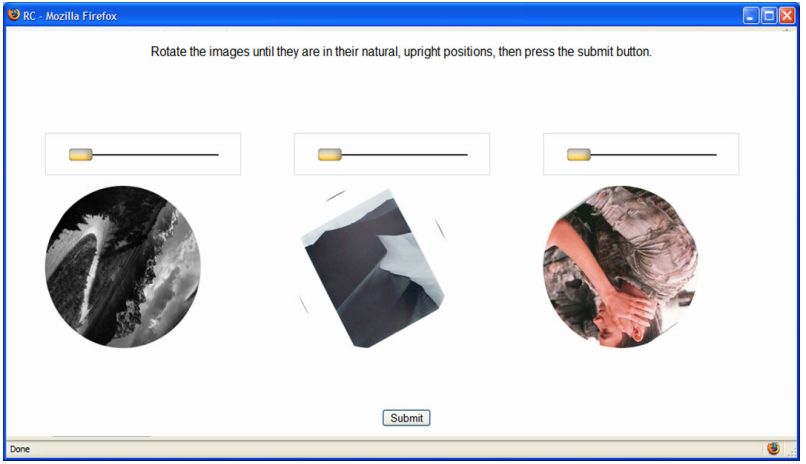}  & 2009 & Move the slider to adjust at least three randomly rotated images to their upright orientation\\   
\cmidrule{2-5} 

  & MintEye CAPTCHA \cite{ref5} &  \includegraphics[width=\imgW, height=\imgL]{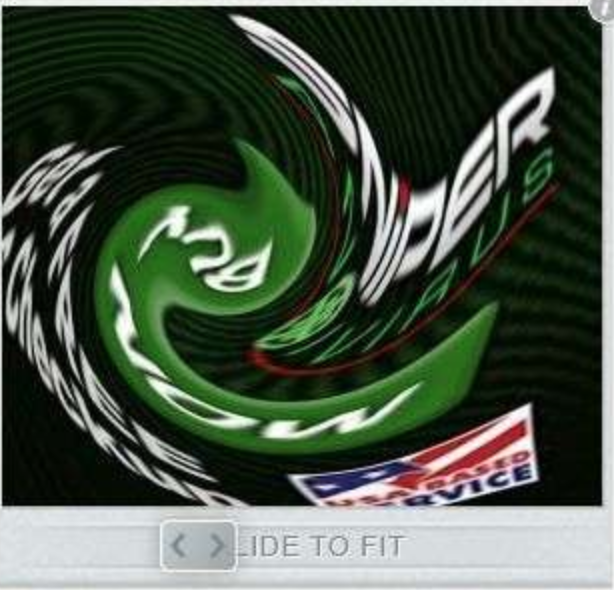} & 2012 & Move the slider until undistorted version of the image appears\\ 
\cmidrule{2-5} 
 & Tencent (Tencent.com) & \includegraphics[width=\imgW, height=\imgL]{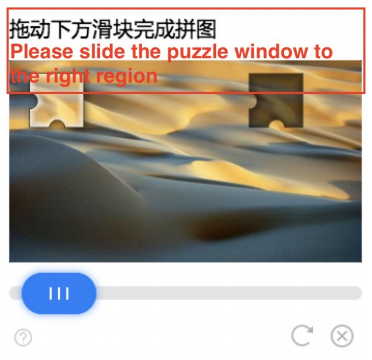} &    & Drag the slider until two puzzle pieces match \\ 
\cmidrule{1-5} 

% Image-based Drag and Drop

 \multirow{5}{*}[-84pt]{Drag and Drop } & Garb CAPTCHA \cite{ref55} & \includegraphics[width=\imgW, height=\imgL]{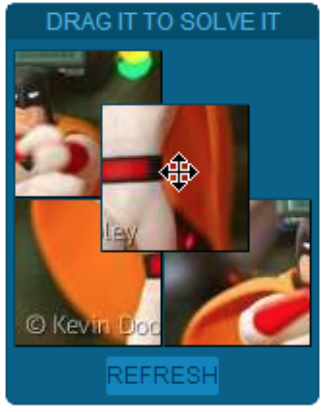} & 2013 & Drag and drop the puzzle pieces to their correct position to reconstruct the original image \\ \cmidrule{2-5} 

 & Capy CAPTCHA \cite{ref53} & \includegraphics[width=\imgW, height=\imgL]{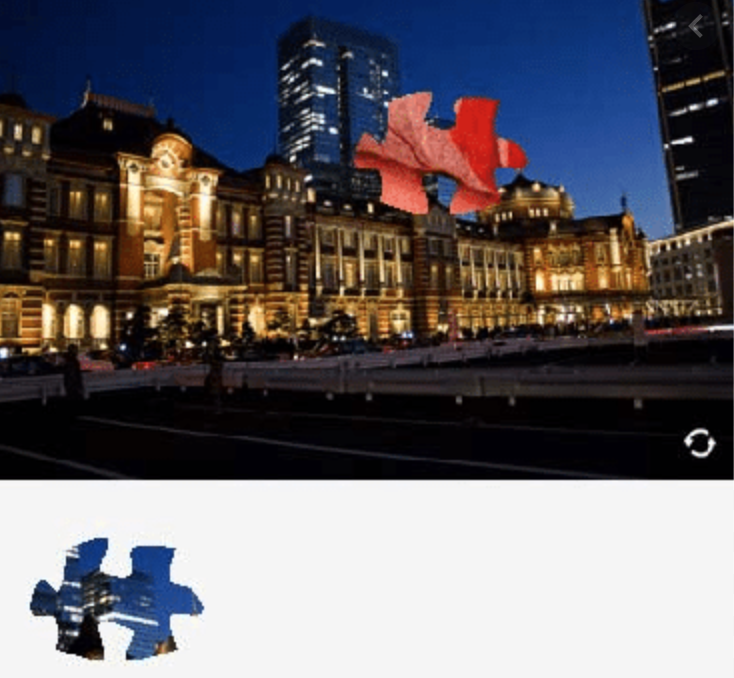} & 2012 & Drag a puzzle piece to complete a jigsaw \\ 
 \cmidrule{2-5} 

 & KeyCAPTCHA \cite{ref54}  & \includegraphics[width=\imgW, height=\imgL]{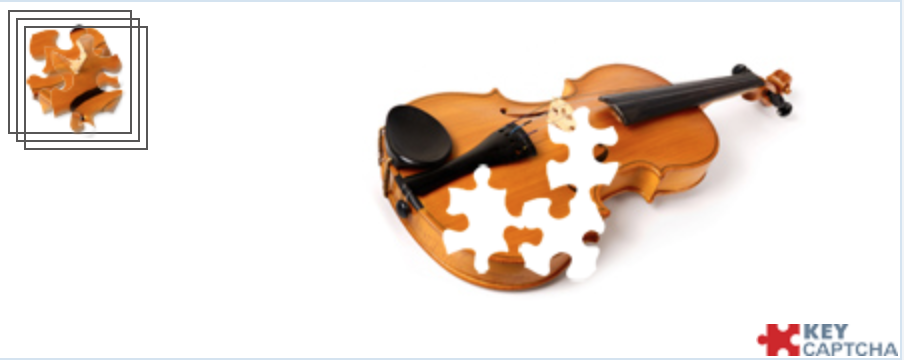}       & 2010 & Drag three puzzle pieces to assemble the image \\ 
\cmidrule{2-5} 

 & Gao et al \cite{ref56} & \includegraphics[width=\imgW, height=\imgL]{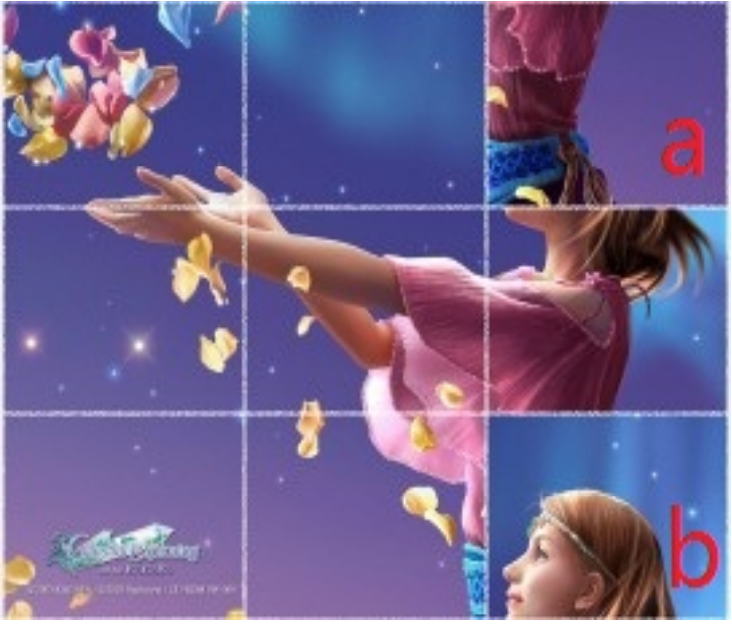}  & 2010 & Identify the two misplaced pieces and swap them \\ 
\cmidrule{2-5} 

 & Hamid Ali et al \cite{ref57} & \includegraphics[width=\imgW, height=\imgL]{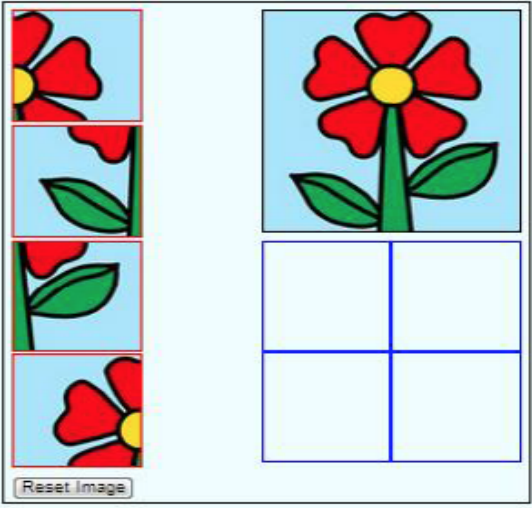} & 2014 & Drag and drop four images to an empty grid following the same order in the reference image \\ 
\cmidrule{1-5} 

% Image-based Selection

  \multirow{9}{*}[-270pt]{Selection} & Asirra \cite{ref38} & \includegraphics[width=\imgW, height=\imgL]{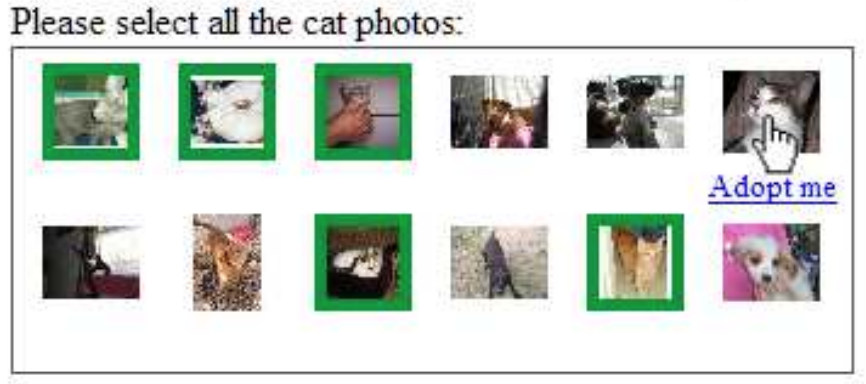} & 2007 & Select cats from a set of 12 images of cats and dogs \\ 
 \cmidrule{2-5} 

  & \multirow{2}{*}[-40pt]{No CAPTCHA reCAPTCHA \cite{ref71}} & \includegraphics[width=\imgW, height=\imgW]{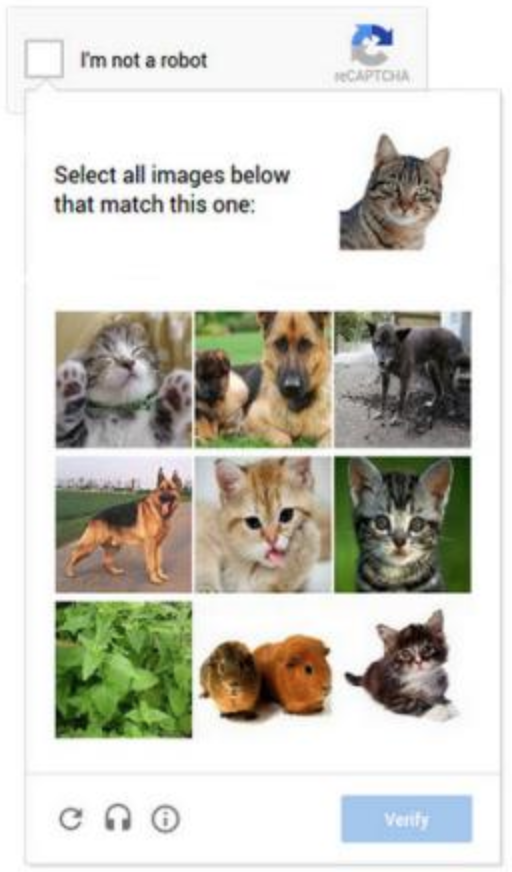} & 2014 & Select images that have the same content described in the challenge with a sample image \\ 
 \cmidrule{3-5}
 
  & & \includegraphics[width=\imgW, height=\imgW]{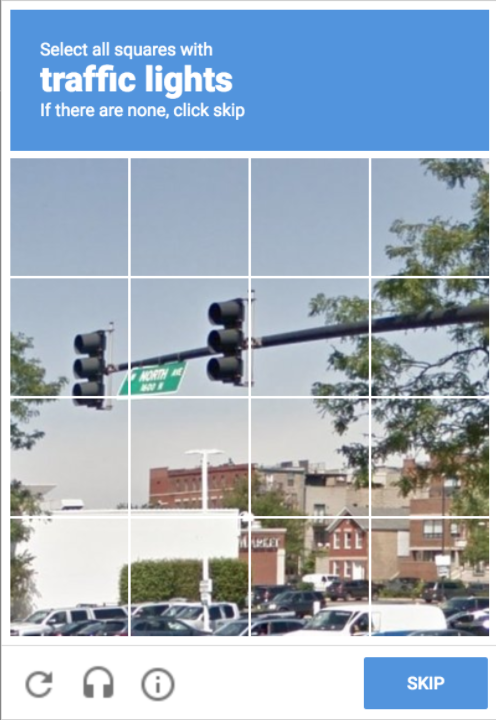} & 2014 & Select all images with street signs, cars, bridges or some specific object \\ 
 \cmidrule{2-5} 
 
 & Facebook image CAPTCHA & \includegraphics[width=\imgW, height=\imgW]{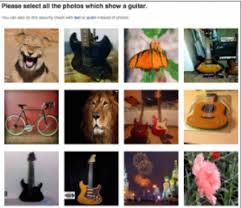} &  & Select the images that correspond to a hint from twelve images with different content \\ 
 \cmidrule{2-5}
 
  & HumanAuth  \cite{ref70} & \includegraphics[width=\imgW, height=\imgL]{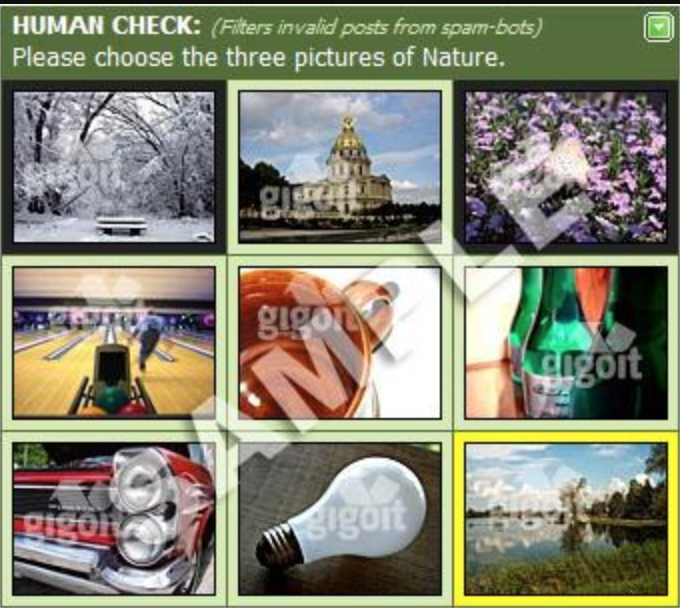} & 2006 & Select images with natural contents \\ 
 \cmidrule{2-5} 

  & SEMAGE \cite{ref39} & \includegraphics[width=\imgW, height=\imgL]{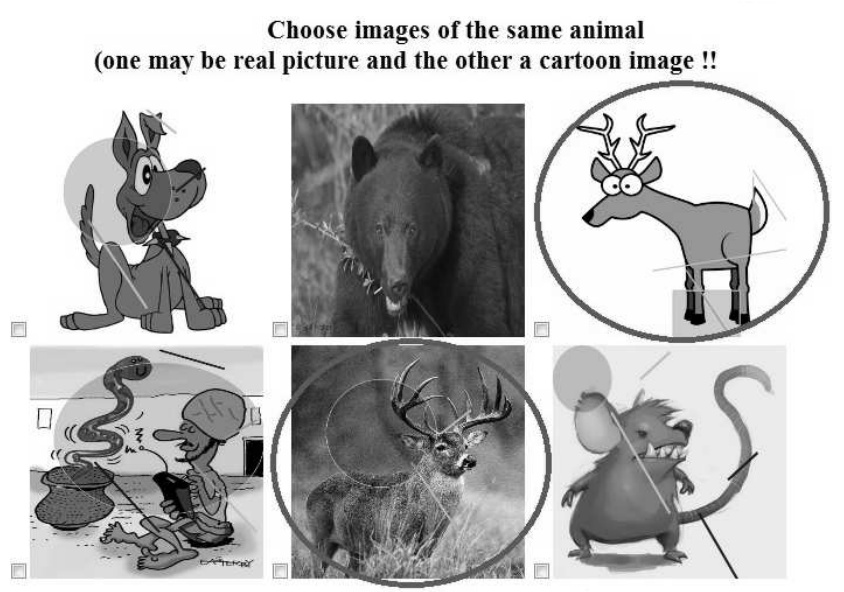} & 2011 & Select semantically related images from a set of images \\ 
 \cmidrule{2-5} 

  & AVATAR \cite{ref43} & \includegraphics[width=\imgW, height=\imgL]{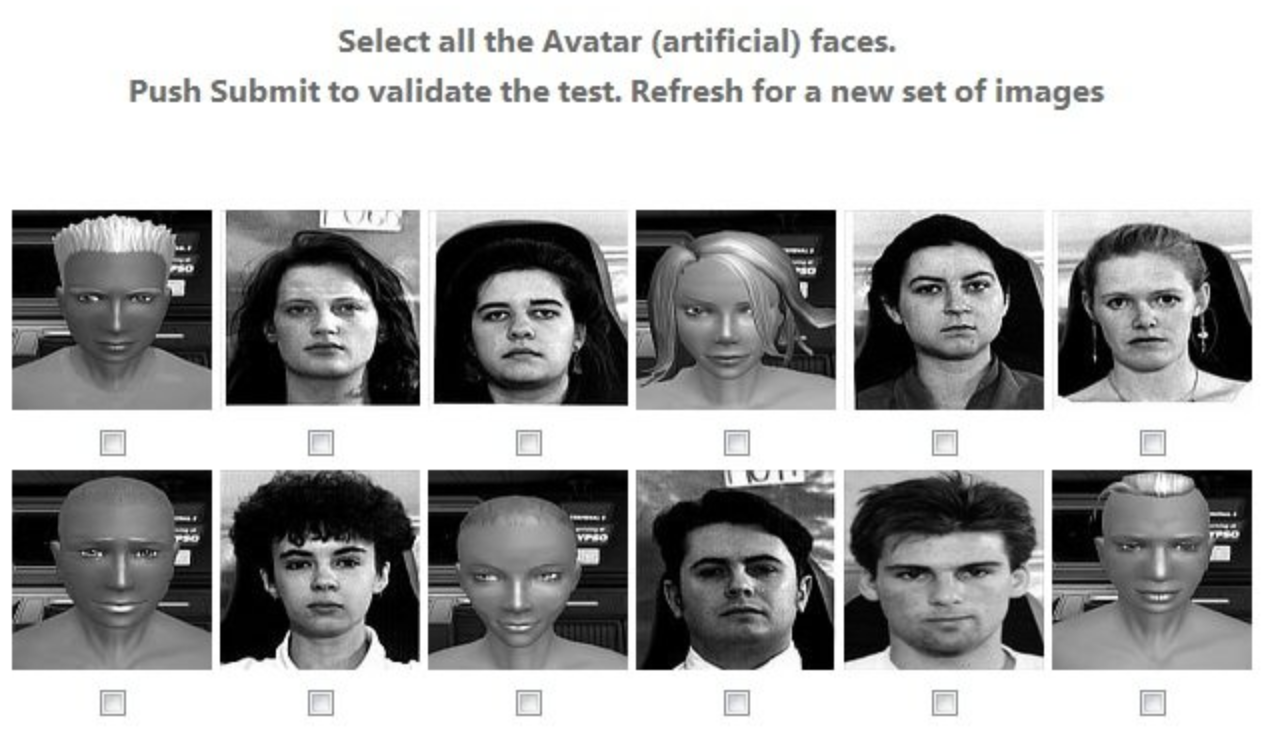} & 2012 & Select avatar faces from a set of 12 images composed of a mix of human and avatar faces\\ 
 \cmidrule{2-5}
 
 & FR-CAPTCHA \cite{ref46} & \includegraphics[width=\imgW, height=\imgL]{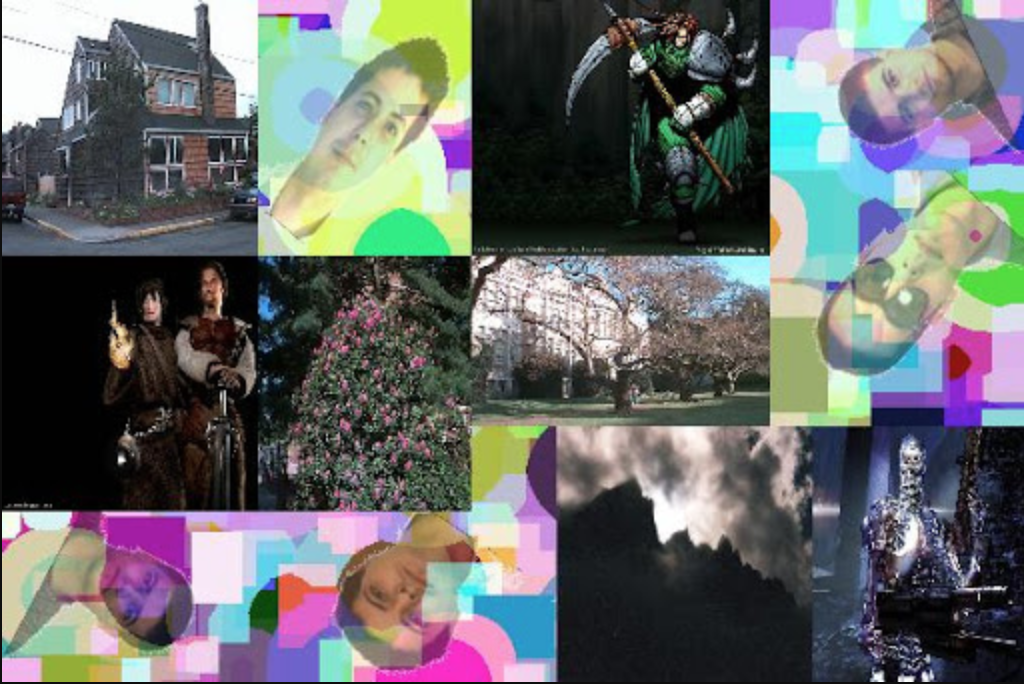} & 2014 & Select real human faces distorted among nonhuman face images \\ 
\cmidrule{2-5} 

 & FaceDCAPTCHA \cite{ref45} & \includegraphics[width=\imgW, height=\imgL]{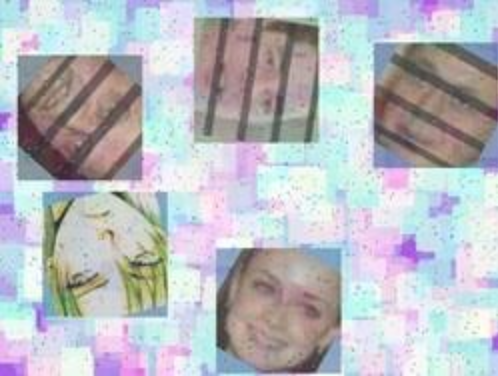} & 2014 & Select two face images of the same person \\ 
\cmidrule{1-5}
% Image-based Drawing

  \multirow{3}{*}[-56pt]{Drawing} & VAPTCHA \cite{ref68}  &    \includegraphics[width=\imgW, height=\imgL]{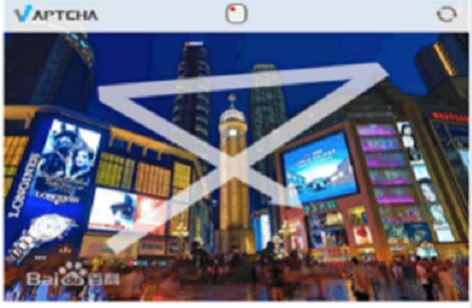}  & 2018 & Draw an resemblant trajectory to match the reference trajectory \\ 
 \cmidrule{2-5} 

 & Drawing CAPTCHA \cite{ref62} &  \includegraphics[width=\imgW, height=\imgL]{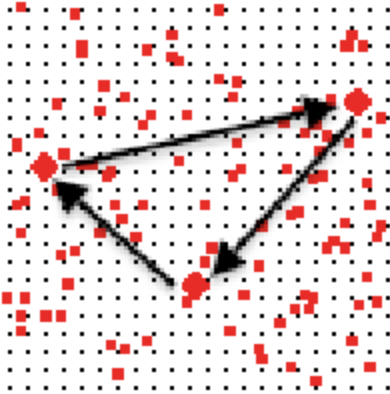} & 2006 & Connect a specific dots to each other \\
\cmidrule{2-5} 

  & MotionCAPTCHA \cite{ref69} & \includegraphics[width=\imgW, height=\imgL]{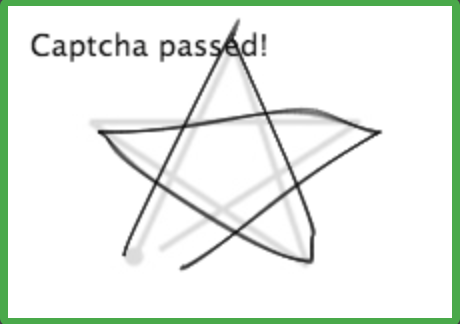} & 2011 & Draw a shape displayed in a box \\
\cmidrule{1-5}

% Image-based interactive

  \multirow{3}{*}{Interactive} &  CAPTCHAStar & \includegraphics[width=\imgW, height=\imgL]{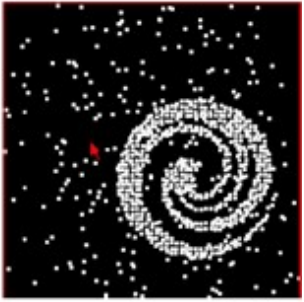} & 2015 & Move the cursor until forming a recognizable shape \\ 
 \cmidrule{2-5} 

 & Cursor CAPTCHA \cite{ref66} &  \includegraphics[width=\imgW, height=\imgL]{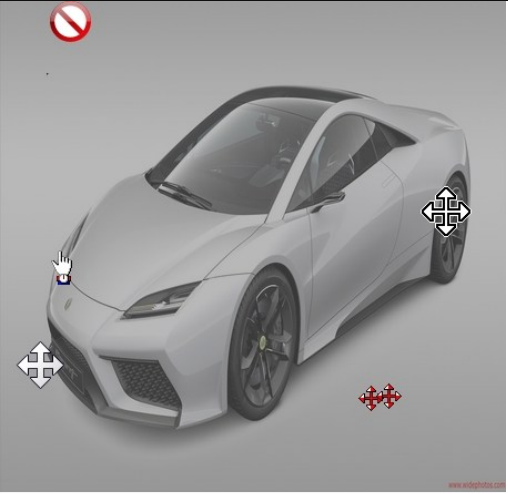} & 2013 & Overlap the cursor on the identical object placed  in a random generated image \\ 
 \cmidrule{2-5} 

  & Noise CAPTCHA \cite{ref65} & \includegraphics[width=\imgW, height=\imgL]{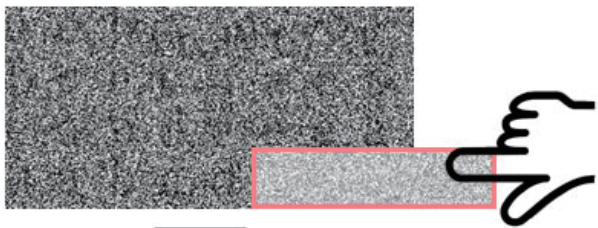} & 2012 & Move a small noisy image over a large noisy image until a hidden message or object appears \\ 
%\cmidrule{2-6}

\midrule

\caption{A taxonomy of image-based CAPTCHAs }
\label{table:imageCAPTCHA}

\end{longtable}
}